\PassOptionsToPackage{pdfpagelabels=false}{hyperref}
\documentclass[useAMS,fleqn,usenatbib]{mnras}
\pdfoutput=1

\usepackage{newtxtext,newtxmath}
\usepackage[T1]{fontenc}
\usepackage{ae,aecompl}

\usepackage{graphicx}   
\usepackage{amsmath}    

\usepackage{xcolor}

\usepackage{subcaption} 
\title[LRG clustering with photo-$z$'s]{The Clustering of DESI-like Luminous Red Galaxies Using Photometric Redshifts}

\author[R. Zhou et al.]{Rongpu Zhou,$^{1, 2, 3}$\thanks{E-mail: rongpuzhou@lbl.gov}
Jeffrey A. Newman,$^{1, 2}$
Yao-Yuan Mao,$^{1, 2, 4}$\thanks{NHFP Einstein Fellow}
Aaron Meisner,$^{5}$\newauthor
John Moustakas,$^{6}$
Adam D. Myers,$^{7}$
Abhishek Prakash,$^{8}$
Andrew R. Zentner,$^{1, 2}$\newauthor
David Brooks,$^{9}$
Yutong Duan,$^{10, 3}$
Martin Landriau,$^{3}$
Michael E. Levi,$^{3}$\newauthor
Francisco Prada$^{11}$
and Gregory Tarle$^{12}$
\\\\\\
$^{1}$Department of Physics and Astronomy, University of Pittsburgh, 3941 O'Hara Street, Pittsburgh, PA 15260, USA\\
$^{2}$Pittsburgh Particle physics, Astrophysics, and Cosmology Center (PITT PACC)\\
$^{3}$Lawrence Berkeley National Laboratory, 1 Cyclotron Road, Berkeley, CA 94720, USA\\
$^{4}$Department of Physics and Astronomy, Rutgers, The State University of New Jersey, Piscataway, NJ 08854, USA\\
$^{5}$NSF's National Optical-Infrared Astronomy Research Laboratory, 950 N. Cherry Avenue, Tucson, AZ 85719, USA\\
$^{6}$Department of Physics and Astronomy, Siena College, 515 Loudon Road, Loudonville, NY 12211\\
$^{7}$Department of Physics \& Astronomy, University  of Wyoming, 1000 E. University, Dept.~3905, Laramie, WY 82071, USA\\
$^{8}$Infrared Processing and Analysis Center, California Institute of Technology, MC 100-22, Pasadena, CA 91125, USA\\
$^{9}$Department of Physics \& Astronomy, University College London, Gower Street, London, WC1E 6BT, UK\\
$^{10}$Physics Department, Boston University, 590 Commonwealth Avenue, Boston, MA 02215, USA\\
$^{11}$Instituto de Astrof\'\i sica de Andaluc\'\i a (CSIC), Glorieta de la Astronom\'\i a, s/n, E-18008 Granada, Spain\\
$^{12}$Department of Physics, University of Michigan, Ann Arbor, MI 48109, USA\\
}

\date{Accepted XXX. Received YYY; in original form ZZZ}

\pubyear{2021}

\begin{document}
\label{firstpage}
\pagerange{\pageref{firstpage}--\pageref{lastpage}}
\maketitle

\begin{abstract}
We present measurements of the redshift-dependent clustering of a DESI-like luminous red galaxy (LRG) sample selected from the Legacy Survey imaging dataset, and use the halo occupation distribution (HOD) framework to fit the clustering signal. The photometric LRG sample in this study contains 2.7 million objects over the redshift range of $0.4 < z < 0.9$ over 5655 deg$^2$. We have developed new photometric redshift (photo-$z$) estimates using the Legacy Survey DECam and WISE photometry, with $\sigma_{\mathrm{NMAD}} = 0.02$ precision for LRGs. We compute the projected correlation function using new methods that maximize signal-to-noise ratio while incorporating redshift uncertainties.  We present a novel algorithm for dividing irregular survey geometries into equal-area patches for jackknife resampling. For a five-parameter HOD model fit using the MultiDark halo catalog, we find that there is little evolution in HOD parameters except at the highest redshifts. The inferred large-scale structure bias is largely consistent with constant clustering amplitude over time. In an appendix, we explore limitations of Markov chain Monte Carlo fitting using stochastic likelihood estimates resulting from applying HOD methods to N-body catalogs, and present a new technique for finding best-fit parameters in this situation.  Accompanying this paper we have released the Photometric Redshifts for the Legacy Surveys (PRLS) catalog of photo-$z$'s obtained by applying the methods used in this work to the full Legacy Survey Data Release 8 dataset. This catalog provides accurate photometric redshifts for objects with $z < 21$ over more than 16,000 deg$^2$ of sky.
\end{abstract}




\section{Introduction}


Galaxy surveys over the past two decades, such as the Sloan Digital Sky Survey (SDSS, \citealt{gunn_sloan_1998}, \citealt{york_sloan_2000}), Baryon Oscillation Spectroscopic Survey (BOSS) \citep{dawson_baryon_2013} and DEEP2 \citep{newman_deep2_2013}, have enabled remarkable advancement in our understanding of the Universe, and more recent and on-going surveys such as the Dark Energy Survey (DES, \citealt{thedarkenergysurveycollaboration_dark_2005}) and the Hyper Suprime-Cam survey \citep{aihara_hyper_2018} are providing more stringent constraints on models of cosmology and galaxy evolution.

The Dark Energy Spectroscopic Instrument (DESI, \citealt{desicollaboration_desi_2016, desi_collaboration_desi_2016}) is a next-generation galaxy redshift survey that will produce an unprecedented 3-dimensional map of the Universe and shed light on the expansion history of the Universe and the nature of dark energy. An important class of DESI spectroscopic targets are the luminous red galaxies (LRGs). The high large-scale structure bias of the LRGs make them ideal tracers for the underlying matter distribution, and they have been used to efficiently measure the baryon acoustic oscillation (BAO) signal (e.g., \citealt{eisenstein_detection_2005}, \citealt{alam_clustering_2017}). The DESI survey, with much larger survey volume and higher number density, will enable more accurate BAO measurements.


In this paper we the present small-scale ($\lesssim 20 h^{-1}\mathrm{Mpc}$) galaxy clustering analysis of a set of DESI-like LRGs. Although the small-scale clustering of LRGs from other programs have been studied previously (e.g., \citealt{zheng_halo_2008}, \citealt{white_clustering_2011}, \citealt{zhai_clustering_2017}), these samples are either at lower redshifts or are much sparser than the DESI LRG sample. This paper presents the first detailed study of the clustering of the DESI-like LRGs.

The selection of our LRG sample is motivated by and intended to mimic the DESI LRG selection. Since spectroscopic redshifts are not available yet, we compute accurate photometric redshifts (photo-$z$'s) and their error estimates using imaging in $g, r$ and $z$ bands from Dark Energy Camera Legacy Survey (DECaLS), which is part of the DESI Legacy Imaging Surveys \citep{dey_overview_2019}, and imaging in $W1$ and $W2$ bands from the Wide-field Infrared Survey Explorer (WISE, \citealt{wright_widefield_2010}). We use the photo-$z$'s to measure the projected correlation functions in five redshift bins over redshift range of $0.4<z\lesssim 0.9$. We interpret the results in the halo occupation distribution (HOD) framework, and we incorporate photo-$z$ error estimates and their uncertainties in this analysis.

Although one can expect that better constraints on HOD parameters will be obtained with spectroscopic redshifts from the upcoming DESI data, this work nevertheless provides an important test of the DESI LRG target selection by providing an estimate of the large-scale structure bias which we can compare with the expected value, and the results also enables the construction of accurate mock galaxies catalogs for DESI analysis before and in the early stage of the survey. Moreover, we demonstrate that data from imaging surveys alone can provide powerful constraints on parameters of HOD and other galaxy-halo connections models.


This paper is organized as follows. We describe the data and the LRG sample in section \ref{sec:data}. We describe the photometric redshifts in section \ref{sec:lrg_photo-z}. We describe the clustering measurements in section \ref{sec:measurements} and modeling of the measurements in section \ref{sec:modeling}. We discuss the results and conclude in section \ref{sec:mcmc_results} and \ref{sec:conclusion}.

We make the photo-$z$'s (computed with more recent DR8 data) publicly available\footnote{\url{https://www.legacysurvey.org/dr8/files/\#photometric-redshifts}}. We describe them in Appendix \ref{sec:dr8_pz}.

\section{Data}
\label{sec:data}

We use the publicly available imaging catalogs from DECaLS DR7 for both sample selection and photometric redshift estimation. DECaLS is one of the DESI imaging surveys, and it provides imaging in $g/r/z$ bands with median $5\sigma$ depth of $24.5/24.0/23.0$ for the fiducial DESI galaxy target (galaxy with an exponential disk profile with half-light radius of $0.45''$). The source catalogs are constructed using the software package \textsc{the Tractor} \citep{2016ascl.soft04008L}
 for source detection and photometry, and they also include the forced photometry of the unWISE coadded images \citep{lang_unwise_2014, meisner_unwise_2019} in the $3.4$ micron ($W1$) and $4.6$ micron ($W2$) bands. Some of the data from DES observations are also processed by \textsc{Tractor} and included in DR7, and are used here.

\subsection{Sample selection}

In the final DESI imaging dataset, each object on average is covered by 3 exposures in each of the three optical bands, and to ensure adequate depth and minimize the impact of cosmic rays we require that each object have at least 2 exposures in each optical band. We remove objects contaminated by nearby bright stars by applying masks as described in section \ref{sec:mask}, and we avoid regions of high stellar density by removing the area within $|b|<25.0\degr$ (where $b$ is the Galactic latitude).

To facilitate the process of dividing the footprint for jackknife resampling (see section \ref{sec:jackknife}), we divide the footprint into HEALPix \citep{gorski_healpix_2005} pixels of area $\sim 0.21$ sq. degrees each (corresponding to $N_{\mathrm{side}} =128$). The pixels at the survey boundaries are removed, and then small ``islands'' consisting of fewer than 100 pixels are also removed.

The photometry is corrected for Galactic extinction using the Galactic transmission values in the DR7 catalog. The LRG sample is selected with the following color and magnitude cuts:
\begin{subequations}
\label{eq:lrg_selection}
\begin{align}
    &(z-W1) > 0.8\times(r-z) - 0.6 \label{eq:non-stellar}\\
    &z < 20.41 \label{eq:mag-limit}\\
    &r-z > (z-17.18)/2 \label{eq:sliding-cut}\\
    &r-z > 0.9 \label{eq:low-z1}\\
    &(r-z > 1.15) \ \mathrm{OR}\  (g-r > 1.65) \label{eq:low-z2}
\end{align}
\end{subequations}

\begin{figure*}
    \includegraphics[width=0.95\textwidth]{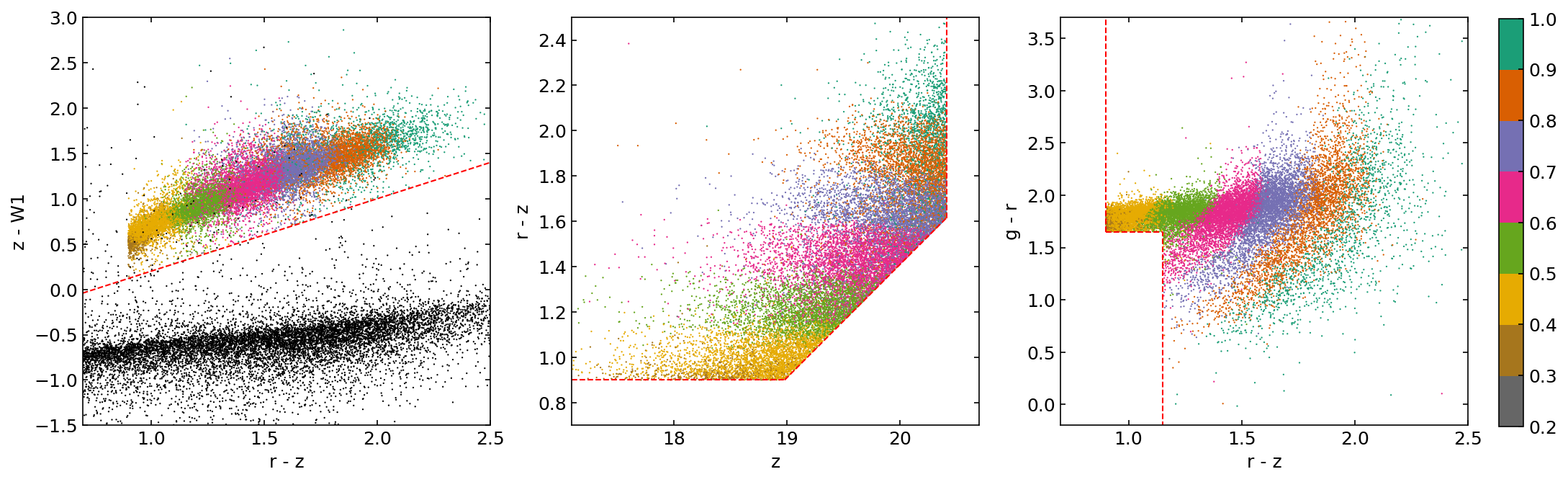}
    \caption{Color-color and color-magnitude diagrams for 20,000 objects randomly selected from the LRG sample. The points are color-coded according to their photometric redshifts (see section \ref{sec:lrg_photo-z}). The dashed lines represent the selection boundaries listed in equations \ref{eq:non-stellar} to \ref{eq:low-z2}. The first panel shows the stellar-rejection cut using $z-W1$ color; in this panel we also plot point sources with $z_\mathrm{mag}<20.41$, the majority of which are stars, in black to show the stellar locus; the stellar-rejection cut removes stars very effectively. The second panel shows the sliding color-magnitude cut and the $z$-band magnitude cut. The third panels shows the cuts that remove low-redshift ($z \lesssim 0.4$) objects.}
    \label{fig:sample_selection}
\end{figure*}

Note that all the magnitudes used in this paper are in the AB system. The cuts are shown in Fig. \ref{fig:sample_selection}. Equation \ref{eq:non-stellar} acts as a stellar-rejection cut, similar to the one presented in \citet{prakash_luminous_2015}.  It is shown in the left panel of Fig. \ref{fig:sample_selection}. This selection relies on the fact that galaxies with old stellar populations have spectra which exhibit a peak at $1.6$ micron (sometimes referred to as the ``1.6 micron bump''); at higher redshift this bump causes an increased flux in the WISE W1 ($3.4$ micron) band. Therefore we can easily separate stars from redshifted galaxies with the $z-W1$ color. This cut allows us to achieve less than 1\% stellar contamination in our sample, as verified using the much deeper and better-seeing HSC data \citep{aihara_first_2018}.

Equation \ref{eq:mag-limit} is an apparent magnitude limit for the sample, which is similar to that expected for DESI LRGs (cf. \citealt{desi_collaboration_desi_2016}). Equation \ref{eq:sliding-cut} is a ``sliding'' color-magnitude cut that imposes a redshift-dependent luminosity threshold on the sample, selecting the most luminous galaxies across the redshift range. Therefore, the sliding cut determines the number density and the shape of the redshift distribution. The magnitude limit and sliding cut are shown in the middle panel of Fig. \ref{fig:sample_selection}.

This sliding cut is combined with the cuts in equations \ref{eq:low-z1} and \ref{eq:low-z2}, which are shown in the right panel of Fig. \ref{fig:sample_selection}, to remove low-redshift ($z \lesssim 0.4$) galaxies and select intrinsically red galaxies. These selection cuts yield roughly uniform comoving number density in the redshift range of $0.4<z \lesssim 0.8$.

\begin{figure*}
    \includegraphics[width=0.99\textwidth]{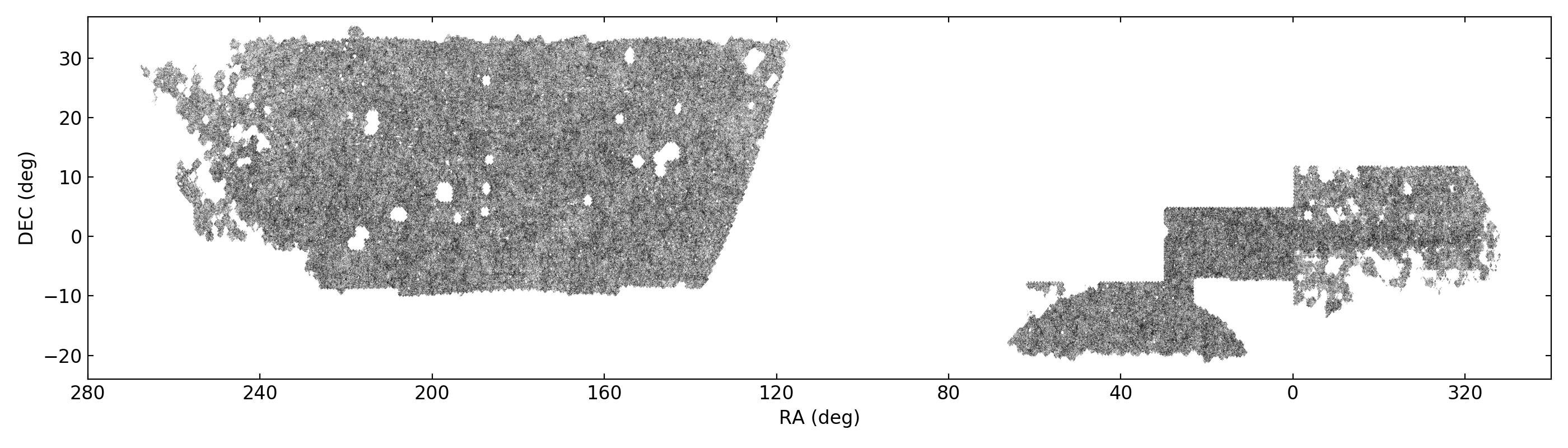}
    \caption{Sky coverage of the final LRG sample that is used in the clustering analysis. The grayscale represents the surface density. The ``holes'' in the NGC footprint and parts of the SGC footprint have been removed from the sample due to contamination from very bright stars or other known imaging artifacts.}
    \label{fig:sky}
\end{figure*}


\subsection{Bright star masks}
\label{sec:mask}
Objects near bright stars are likely to have inaccurate flux measurements due to contamination. Such contamination causes many objects to be selected as LRGs although their true fluxes do not satisfy the selection cuts. The inaccuracies in the WISE PSF modeling lead to overestimated fluxes for objects near even moderately bright stars, and this is a significant source of imaging systematics for the LRG sample. Another issue is that extremely bright stars produce imaging artifacts such as ghosts and diffraction spikes which are poorly modeled. Such artifacts in the optical imaging, which is used for source detection, cause spurious sources as well as inaccurate flux measurements. Therefore we apply masks surrounding the positions of bright stars for both optical and WISE imaging when constructing the LRG sample.

Three different sets of masks are used. First, we use the ``bright-star-in-blob`` column in the catalog. As defined in the DR7 catalog, a blob is a ``contiguous region of pixels above a detection threshold and neighboring pixels''
\footnote{\url{https://www.legacysurvey.org/dr7/description/\#glossary}}
, and an object is flagged if it is in the same blob as a Tycho-2 star \citep{2000A&A...355L..27H}. Second, we use the unWISE masks described in the appendix of \citet{meisner_unwise_2019} to remove areas around bright stars selected from AllWISE \citep{2013wise.rept....1C} and 2MASS \citep{skrutskie_two_2006}. Third, we develop and apply a set of WISE masks that include fainter AllWISE stars that are not in the unWISE masks but still cause significant contamination. The third set of masks is specifically optimized for the LRGs. More about the WISE masks can be found in Appendix \ref{sec:app_masks}. The three sets of masks combined remove ${\sim}\,12\%$ of the objects from the LRG sample but only ${\sim}\,4\%$ of the total area; it is clear that most of the sources masked do not truly belong in the sample.

In addition to the bright star masks, we also remove regions that are affected by very bright stars or other imaging artifacts. We identify such regions by examining areas with a high density of LRGs that have very large ($>0.05$) estimated photo-$z$ errors. Since real LRGs typically have much smaller photo-$z$ errors, these objects are mostly either spurious sources or real sources that are not included in the photo-$z$ training such as stars and quasars; however, only the spurious sources are likely to be highly concentrated on the sky (e.g., around very bright stars). Such concentrations are identified efficiently with the DBSCAN cluster analysis routine in scikit-learn \citep{pedregosa_scikit-learn_2011} and the corresponding HEALPix pixels are flagged as bad regions.

The final LRG sample has 2.74 million objects spanning 5655 sq. degrees. Fig. \ref{fig:sky} shows the sky distribution.

\subsection{Randoms}
The calculation of correlation functions requires uniformly distributed random points with the same survey geometry as the LRG sample. We use the publicly available random catalog for DECaLS DR7
\footnote{\url{https://www.legacysurvey.org/dr7/files/}}. The same number of exposure requirements, footprint cuts, and bright star masks are applied on the randoms as are used in constructing the LRG sample.

\section{Photometric redshifts}
\label{sec:lrg_photo-z}

We compute photometric redshifts using the random forest regression method \citep{Breiman2001}, a machine learning (ML) algorithm based on decision trees. For our dataset, the ML methods have several advantages over template-fitting methods: there are abundant spectroscopic observations of galaxies covering the magnitude and color space of the LRG sample that can be used for training, and in this regime ML methods consistently out-perform template-fitting methods; ML methods can trivially incorporate non-photometry information such as galaxy shapes, which we exploit here; and ML methods do not require physical and representative SED templates, which are not trivial to obtain especially for the wavelength range of the WISE pass-bands. Among the numerous ML methods, random forest provides good performance and is very computationally efficient, so we use it here.

\subsection{Imaging data}

We include $r$-band magnitude as well as $g-r$, $r-z$, $z-W1$ and $W1-W2$ colors as inputs. The photometry has been corrected for Galactic extinction. \citet{soo_morpho-z_2018} showed that while morphological information only provides mild improvements in photo-$z$ accuracy when full $ugriz$ photometry is available, the improvement is substantial when only $grz$ photometry is available. Motivated by that result, we include as inputs three morphological parameters: half-light radius, axis ratio (ratio between semi-minor and semi-major axes), and a ``model weight'' that characterizes whether a galaxy is better fit by an exponential profile or a de Vaucouleurs, similar to the definition in \citet{soo_morpho-z_2018}:
\begin{equation}
\label{eq:nolabel9}
    p = \frac{d\chi^2_{deV}}{d\chi^2_{deV} + d\chi^2_{exp}};
\end{equation}
where $d\chi^2$ is the difference in $\chi^2$ between the model fit and no source. The inclusion of the three morphological parameters significantly reduces the photo-$z$ scatter for bright galaxies: the normalized median absolute deviation, $\sigma_{\mathrm{NMAD}}$, is reduced by $15\%$ for SDSS (not including BOSS or eBOSS) compared to photo-$z$'s computed without using the morphological parameters; for GAMA galaxies the improvement in $\sigma_{\mathrm{NMAD}}$ is $19\%$ and the 10\% outlier fraction is also reduced by ${\sim}\,30\%$.  The improvement is less significant for fainter galaxies. E.g., VIPERS galaxies have a $6\%$ improvement in $\sigma_{\mathrm{NMAD}}$ and $12\%$ improvement in the outlier fraction. This is expected, because the morphological parameters of fainter galaxies are less well constrained due to the lower S/N, and because at higher redshifts the angular diameter of a galaxy decreases more slowly with redshift, thus making the galaxy size a less useful quantity for redshift estimation.

\subsection{Redshift ``truth'' dataset}
\label{sec:truth}
For machine learning photo-$z$ methods, redshift ``truth'' values are needed for the training process. Various redshift surveys overlap with the DECaLS footprint, and we compile a redshift truth dataset using   spectroscopic and many-band photometric redshifts from ten different surveys.

\subsubsection{2dFLenS}
The 2-degree Field Lensing Survey \citep{blake_2-degree_2016} is a spectroscopic survey performed at the Anglo-Australian Telescope, and observed two galaxy samples: a sample of LRGs, and a magnitude limited ($r<19.5$) galaxy sample. We apply the following quality cuts to this sample:
\begin{equation}
\label{eq:nolabel10}
    (Q==4)\ \mathrm{AND}\ (z>0),
\end{equation}
where $Q$ is the quality flag.

\subsubsection{AGES}
The AGN and Galaxy Evolution Survey \citep{kochanek_ages_2012} is a spectroscopic survey performed with the Hectospec instrument at the MMT telescope. Targets were selected with optical and IR imaging down to $\mathrm{I} \simeq 20$ (Vega magnitude). Only objects from the galaxy targets are used, and we required $z>0$.

\subsubsection{COSMOS2015 photo-$z$'s}
The COSMOS2015 catalog \citep{laigle_cosmos2015_2016} is a photometric redshift catalog covering the 2 $\mathrm{deg}^2$ COSMOS field. We include the COSMOS photo-$z$’s because there are few spec-z samples with comparable depth and redshift coverage, and adequate coverage in depth and redshift in the training set is necessary for machine-learning-based methods to work. To minimize the risk of photo-$z$ outliers in COSMOS impacting our training, we apply a set of stringent quality cuts motivated by those which were applied in \citet{tanaka_photometric_2018}:

\begin{enumerate}
    \item FLAG\_PETER is false (no bad photometry)
    \item $\mathrm{TYPE==0}$ (only galaxies)
    \item $(\mathrm{ZPDF\_H68}-\mathrm{ZPDF\_L68})/(1+z)<0.02$    (limit photo-$z$ errors to 1\%)
    \item $(\mathrm{CHI2\_BEST}<\mathrm{CHIS})\ \mathrm{AND}\  (\mathrm{CHI2\_BEST}/\mathrm{NBFILT}<5)$ (fits are reasonable and better than stellar alternatives)
    \item $\mathrm{ZP}\_2<0$ (no secondary peaks)
    \item $\mathrm{MASS\_MED}>7.5$ (stellar mass recovery successful)
    \item $\mathrm{DEC}>1.46$ (removing some apparent imaging artifacts near the lower boundary)
    \item $z>0.006$ (remove the lowest redshift bin)
    \item $z<3$ (redshift upper limit)
\end{enumerate}

We find that if we exclude the COSMOS photo-$z$’s from the training sample, the change in photo-$z$ scatter is negligible (<0.1\% of the original value) for spec-$z$ objects with $z_{\mathrm{mag}}<20.5$; for fainter spec-$z$ objects with $20.5<z_{\mathrm{mag}}<21.5$, the photo-$z$ scatter increases by 1\% if excluding COSMOS photo-$z$’s from the training sample. We also have checked for possible photo-$z$ biases that might be introduced by the COSMOS photo-$z$'s by comparing the mean offsets in bins of $z_{\mathrm{spec}}$ (e.g., the red line in the lower panel of Fig. \ref{fig:dr8_north_pz_bright}) with and without COSMOS photo-$z$'s in the training sample. We find no noticeable difference in the mean offsets for $z_{\mathrm{mag}}<20.5$ objects and reduced offsets for $z_{\mathrm{spec}}<0.5$ objects with $20.5<z_{\mathrm{mag}}<21.0$ when COSMOS photo-$z$'s are included in the training sample (testing only with objects with fully spectroscopic redshift measurements).

\subsubsection{DEEP2}
DEEP2 \citep{newman_deep2_2013} is a spectroscopic redshift survey performed with the DEIMOS instrument on the Keck 2 telescope. Targets were selected down to $R_{AB}=24.1$ with color cuts to exclude $z<0.7$ galaxies in 3 of the 4 fields surveyed. We require that $z>0$ and the quality flag $\mathrm{ZQUALITY}\geq3$.

\subsubsection{GAMA DR3}
The Galaxy And Mass Assembly (GAMA) survey \citep{baldry_galaxy_2018} is a spectroscopic survey performed at the Anglo-Australian Telescope with magnitude limited target selection down to $r \simeq 20$. We require that the quality flag $nQ==4$ and $z>0.002$.

\subsubsection{OzDES}
OzDES is a spectroscopic follow-up survey \citep{childress_ozdes_2017} of the DES supernova fields performed at the Anglo-Australian Telescope. Various types of targets were selected, such as supernova hosts, AGNs and LRGs. For our purposes, we only use objects that were targeted as ``LRG'', ``bright galaxy'', ``ELG'', ``photo-$z$'', ``RedMaGiC'' or ``cluster galaxy''. We also require that the quality flag $Q==4$ and $z>0$.

\subsubsection{SDSS DR14}
We use spectroscopic redshifts from SDSS DR14 \citep{abolfathi_fourteenth_2018}, including the SDSS Main Galaxy Sample \citep{strauss_spectroscopic_2002}, Baryon Oscillation Spectroscopic Survey (BOSS) sample \citep{dawson_baryon_2013}, and the Extended Baryon Oscillation Spectroscopic Survey (eBOSS) sample \citep{dawson_sdss-iv_2016}. To select objects with accurate photo-$z$'s, we apply the following quality cuts:

\begin{enumerate}
    \item $\mathrm{ZWARNING}==0$ (no known problems)
    \item $\mathrm{CLASS}==\mathrm{GALAXY}$ (classified as galaxy)
    \item $z>0.0003$ (remove spurious galaxies at very low redshift)
\end{enumerate}

\subsubsection{VIPERS}
The VIMOS Public Extragalactic Redshift Survey \citep{scodeggio_vimos_2018} is a spectroscopic survey performed at the ESO VLT. The sample is magnitude limited to $i=22.5$ with color cuts to exclude $z<0.5$ galaxies. To ensure redshift quality, we require $3.4 \leq \mathrm{zflg} < 5$ where $\mathrm{zflg}$ is the quality flag, and $z>0$.

\subsubsection{VVDS}
The VIMOS VLT Deep Survey \citep{le_fevre_vimos_2013} is a spectroscopic survey performed at the ESO VLT. The sample is i-band selected, down to $i=22.5$ in the wide field and $i \simeq 24$ in the deeper field. We require that the quality flag $\mathrm{ZFLAGS}==4$ and $z>0$.

\subsubsection{WiggleZ}
The WiggleZ Dark Energy Survey \citep{parkinson_wigglez_2012} is a spectroscopic survey performed at the Anglo-Australian Telescope that aims to measure the baryon acoustic oscillations (BAO) signal with emission-line galaxies. We require that the quality flag $Q==4\ \mathrm{or}\ 5$, and we also require $z>0$.

\subsection{Combined truth dataset and downsampling}
\begin{figure}
    \includegraphics[width=0.9\columnwidth]{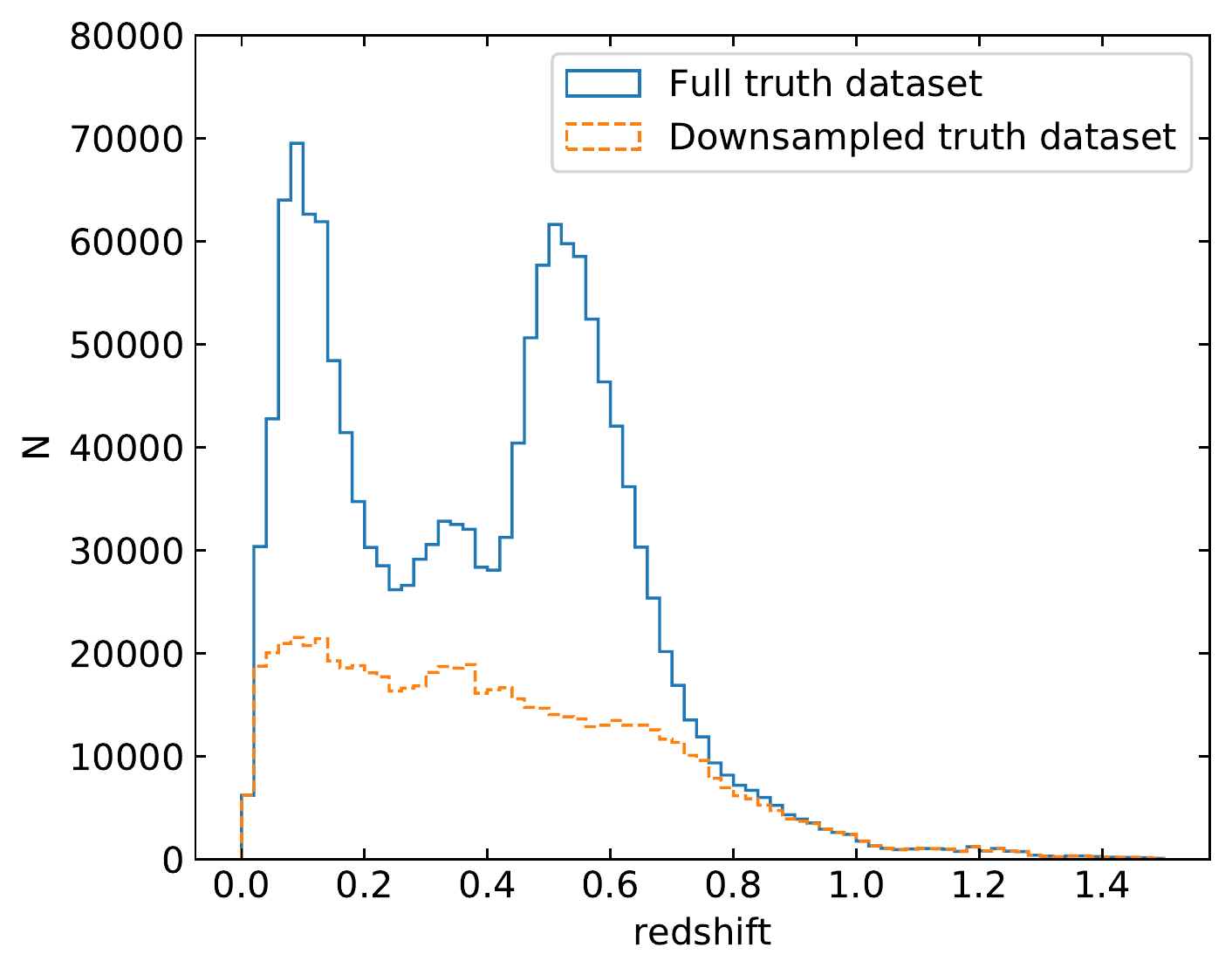}
    \caption{Redshift distribution of the redshift truth dataset. N(z) is the total number of objects in each $\Delta z=0.02$ bin. The peaks at $z=0.1$ and $z=0.5$ in the full set are attributed to the SDSS Main Galaxy Sample and BOSS, respectively, both of which are downsampled significantly to avoid biasing the output photo-$z$'s to favor these redshifts.}
    \label{fig:truth_redshift_distribution}
\end{figure}

The redshift catalogs are cross-matched to DECaLS with a search radius of $1''$. Fig. \ref{fig:truth_redshift_distribution} shows the redshift distribution of the cross-matched redshift truth dataset. Table \ref{tab:truth_catalogs} lists the number of objects from each survey.

\begin{table}
    \centering
    \caption{Number of objects from each redshift survey that are cross-matched to DECaLS.}
    \label{tab:truth_catalogs}
    \begin{tabular}{lll}
        \hline
        Survey & Full dataset & Downsampled dataset\\
        \hline
        BOSS        &  678370  &  224345  \\
        SDSS        &  449386  &  186666  \\
        WiggleZ     &  122907  &  47334   \\
        GAMA        &  109790  &  55990   \\
        COSMOS2015  &  53973   &  53972   \\
        VIPERS      &  44175   &  44175   \\
        eBOSS       &  23549   &  23549   \\
        DEEP2       &  15994   &  15994   \\
        AGES        &  11235   &  11235   \\
        2dFLenS     &  8102    &  8102    \\
        VVDS        &  5490    &  5490    \\
        OzDES       &  1407    &  1407    \\
        \hline
    \end{tabular}
\end{table}

Most of the truth objects are from four surveys: SDSS, BOSS, GAMA and WiggleZ. These surveys either apply specific color selections (BOSS, WiggleZ) or are limited to shallow magnitudes (SDSS Main Galaxy Sample, GAMA). This leads to sharp peaks in the redshift distribution and discontinuities in the density of objects in color/magnitude space. Such a non-uniform training sample can cause systematic biases in the photo-$z$'s; e.g., photo-$z$ algorithms can develop tendency to favor placing objects at redshifts which are over-represented in training set. To make the training sample more uniform and also to speed up computation, objects from these four large surveys are downsampled. This downsampling is based on the object density in the two-dimensional space of $r$-band magnitude (used as a proxy for luminosity at fixed redshift) vs. redshift, with a bin size of $\Delta z=0.01$ and $\Delta r_{\mathrm{mag}}=0.01$. For $r_{\mathrm{mag}}$-redshift bins that have more galaxies than a specific threshold (which are 400, 400, 70 and 20 for SDSS, BOSS, GAMA and WiggleZ, respectively), the objects are randomly downsampled so that the density is reduced to the threshold level. In this way we reduce the overall number of galaxies while preserving a good sampling of galaxies over the full range of luminosity, in particular retaining the rare most luminous galaxies, many of which are LRGs.
The redshift distribution of the downsampled truth dataset is shown in Fig. \ref{fig:truth_redshift_distribution}. Hereafter, we refer to the downsampled truth catalog simply as the truth catalog.

\subsection{Random forest method}
\label{sec:rf}
We compute photo-$z$'s using the random forest regression routine in Scikit-Learn \citep{pedregosa_scikit-learn_2011}. As we have described previously, we use the following eight parameters as input: $r$-band magnitude, $g-r$, $r-z$, $z-W1$ and $W1-W2$ colors, half-light radius, axis ratio and shape probability.

The redshift and magnitude distributions of the photo-$z$ training sample are not uniform due to the various selections of the spectroscopic surveys. In the presence of photometric errors, the gradients in the color and magnitude distributions cause objects in higher density regions in the multi-dimensional color/magnitude space to be scattered into lower density regions. Therefore in the neighborhood of each point in the color/magnitude space, it is more likely to find objects from higher density locations, and since colors and magnitudes are correlated with redshift (which is why photo-$z$ algorithms work), this causes the photo-$z$ estimates to be biased towards the redshifts of objects in higher density regions in the parameter space. 
Such bias is particularly significant at the high-redshift end of the redshift distribution, where the photo-$z$ estimates are consistently biased low. To mitigate this bias, we assign weights to each training object based on its spectroscopic redshift (or photometric redshift, in the case of COSMOS): the weight is proportional to the inverse of the number of training objects at that redshift (with a cut-off value to prevent excessively large weights). As a result, objects at very low or very high redshifts are assigned larger weights than other objects. These weights are incorporated into our random forest calculation by requiring a minimum of 25 objects to split an internal node, and then using the weighted average of the redshifts for all training objects in a leaf as its predicted $z$.

We randomly select 90\% of the truth dataset for training, and reserve the other 10\% for testing purposes. To estimate the photo-$z$ error for each object, we perturb the photometry by adding to the observed flux in each band a random value from Gaussian distribution whose standard deviation is set by the photometric error. This is similar to \citet{kind_tpz_2013}, although in that work the photometry of the training sample, rather than the sample of objects to which the algorithm is applied as here, was perturbed. Ideally we would perturb the morphological parameters as well, but were not able to because the errors on these parameters are either not applicable (for the $d\chi^2$-based ``model weight'') or not provided (for the half-light radius and the axis ratio) in the imaging catalog.

For each of the 50 individual trees in the random forest that we generated, we repeat the perturbation 20 times. The mean and standard deviation of the resulting 1000 ($50\times20$) redshift estimates are used as the photo-$z$ and photo-$z$ error, respectively. Note that the photometric noise is only added when computing the final photo-$z$'s used for clustering analysis; the random forest is trained with the original unperturbed photometry.

\subsection{Photo-$z$ performance for LRGs}
\label{sec:validation}

Here we describe the photo-$z$ performance for our LRG sample. For the photo-$z$ performance of the overall spectroscopic training sample, see Appendix \ref{sec:dr8_pz}.

\begin{figure}
    \includegraphics[width=0.9\columnwidth]{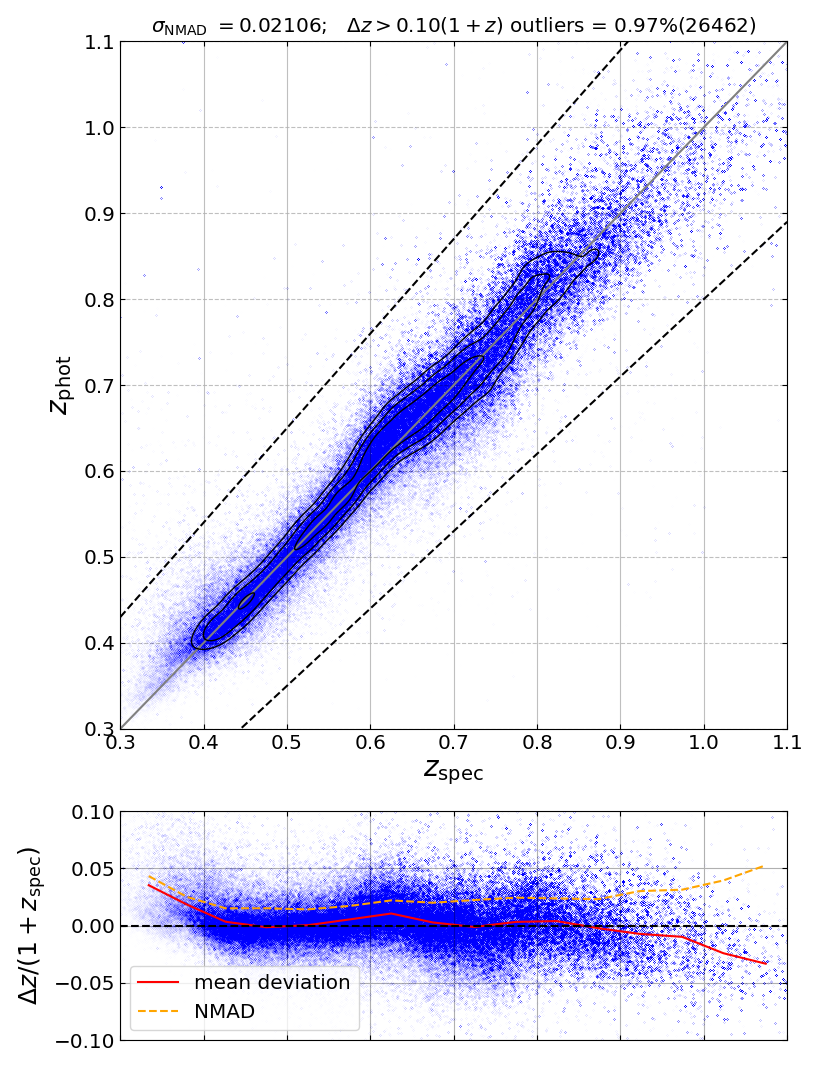}
    \centering
    \caption{Top panel: Photo-$z$ vs. spec-$z$ for truth objects that are weighted to approximate the photo-$z$ performance of the LRG sample. Contours are used to indicate density where the points are densest.  Black dashed lines indicate the boundaries outside which objects are classified as outliers. Lower panel: photo-$z$ offset (in $\Delta z/(1 + z_{\mathrm{spec}}$) vs. spec-$z$; the red solid line and the yellow dashed line are the Hodges-Lehmann mean and $\sigma_{\mathrm{NMAD}}$, respectively, of the photo-$z$ offset in bins of spec-$z$. The photo-$z$'s are mostly well constrained with few outliers for this sample.}
    \label{fig:pz_weighted}
\end{figure}

To assess the photo-$z$ quality of the LRG sample, we cross-match the LRG sample to the truth catalogs in the multi-dimensional space of $r$-band magnitude and $g-r$, $r-z$, $z-W1$ and $W1-W2$ colors; redshift information is not used in the matching. Each LRG is matched to the single nearest neighbor in the truth catalog, and we count the number of LRGs that each truth object is matched to. We use this number as the weight for the photo-$z$ vs. spec-$z$ plot in Fig. \ref{fig:pz_weighted}. 
We quantify the photo-$z$ accuracy using the normalized median absolute deviation (NMAD), defined as $\sigma_{\mathrm{NMAD}} = 1.48 \times \mathrm{median}(|\Delta z|/(1 + z_{\mathrm{spec}}))$ where $\Delta z = z_{\mathrm{phot}} - z_{\mathrm{spec}}$ (we follow the same NMAD definition as in, e.g.,  \citealt{dahlen_critical_2013}). This is a robust estimator of scatter as it is not sensitive to outliers. We also measure the fraction of outliers defined as objects with $|\Delta z|>0.1 \times (1 + z_{\mathrm{spec}})$.
From the weighted spec-$z$ objects, we estimate that the average photo-$z$ scatter $\sigma_{\mathrm{NMAD}}$ for the LRG sample is $0.021$ and the outlier rate is $1\%$. The weighted spec-$z$ objects are also used for estimating the comoving number density which we use as an observable in our HOD modeling (see section \ref{sec:modeling}).

\subsubsection{Photo-$z$ error estimates}
\label{sec:pz_error_validation}
The photo-$z$ error estimation procedure described in \S \ref{sec:rf} is built upon the assumptions that 1) the photo-$z$ probability density functions (PDFs) are Gaussian; 2) photo-$z$ uncertainties are dominated by photometric errors; and 3) the photometric error estimates are accurate. They do not include the effects of incompleteness in the training data or uncertainties in morphological parameters. 

\begin{figure}
    \includegraphics[width=1.03\columnwidth]{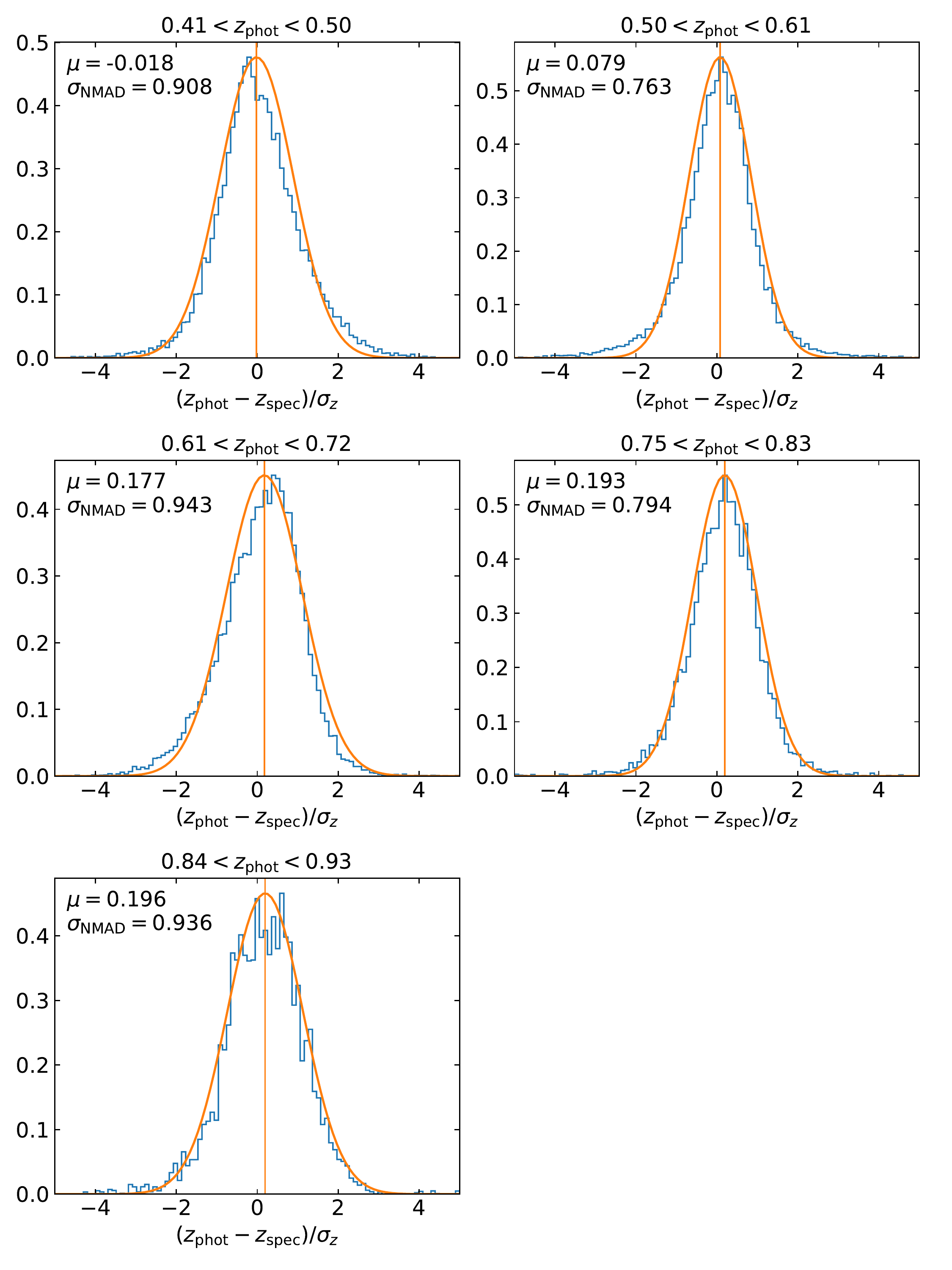}
    \centering
    \caption{The distribution of $(z_{\mathrm{phot}}-z_{\mathrm{spec}})/\sigma_z$, where $\sigma_z$ is the estimated photo-$z$ error, in different redshift bins. Here $\mu$ (vertical line) and $\sigma_\mathrm{NMAD}$ (1-$\sigma$ width of the curve) are the median and the normalized normalized median absolute deviation of the distribution, respectively, and the smooth curves show the corresponding Gaussian distributions. A non-zero $\mu$ value indicates that the photo-$z$'s are biased, and any deviation of $\sigma_\mathrm{NMAD}$ from unity indicates over- or under-estimation of the photo-$z$ errors. The fact that $\sigma_\mathrm{NMAD}$ values are consistently less than unity indicates that the our photo-$z$ errors are over-estimated.}
    \label{fig:pz_error_validation}
\end{figure}

We have validated the photo-$z$ error estimates using objects with spectroscopic redshifts. 
If the photo-$z$ PDFs are indeed Gaussian and the estimated photo-$z$ errors are accurate, the photo-$z$ of each galaxy should differ from its spec-$z$ by a random value $\Delta z$ drawn from a Gaussian distribution with $\sigma$ equal to the estimated photo-$z$ error, i.e., $\Delta z/\sigma_z$, where $\sigma_z$ is the estimated photo-$z$ error for a given object, should be a standard normal distribution with mean 0 and standard deviation 1. As shown in Fig. \ref{fig:pz_error_validation}, we find that $\Delta z/\sigma_z$ is roughly consistent with a Gaussian distribution, but the photo-$z$ errors are generally overestimated by a factor which depends on the redshift range. Because we do not have representative spectroscopic data to estimate this scaling factor accurately, we treat it as nuisance parameter in the HOD modeling (see section \ref{sec:mock}).

\section{Clustering measurements}
\label{sec:measurements}

\subsection{Redshift bins}
\label{sec:redshift_bins}
\begin{figure}
    \includegraphics[width=0.95\columnwidth]{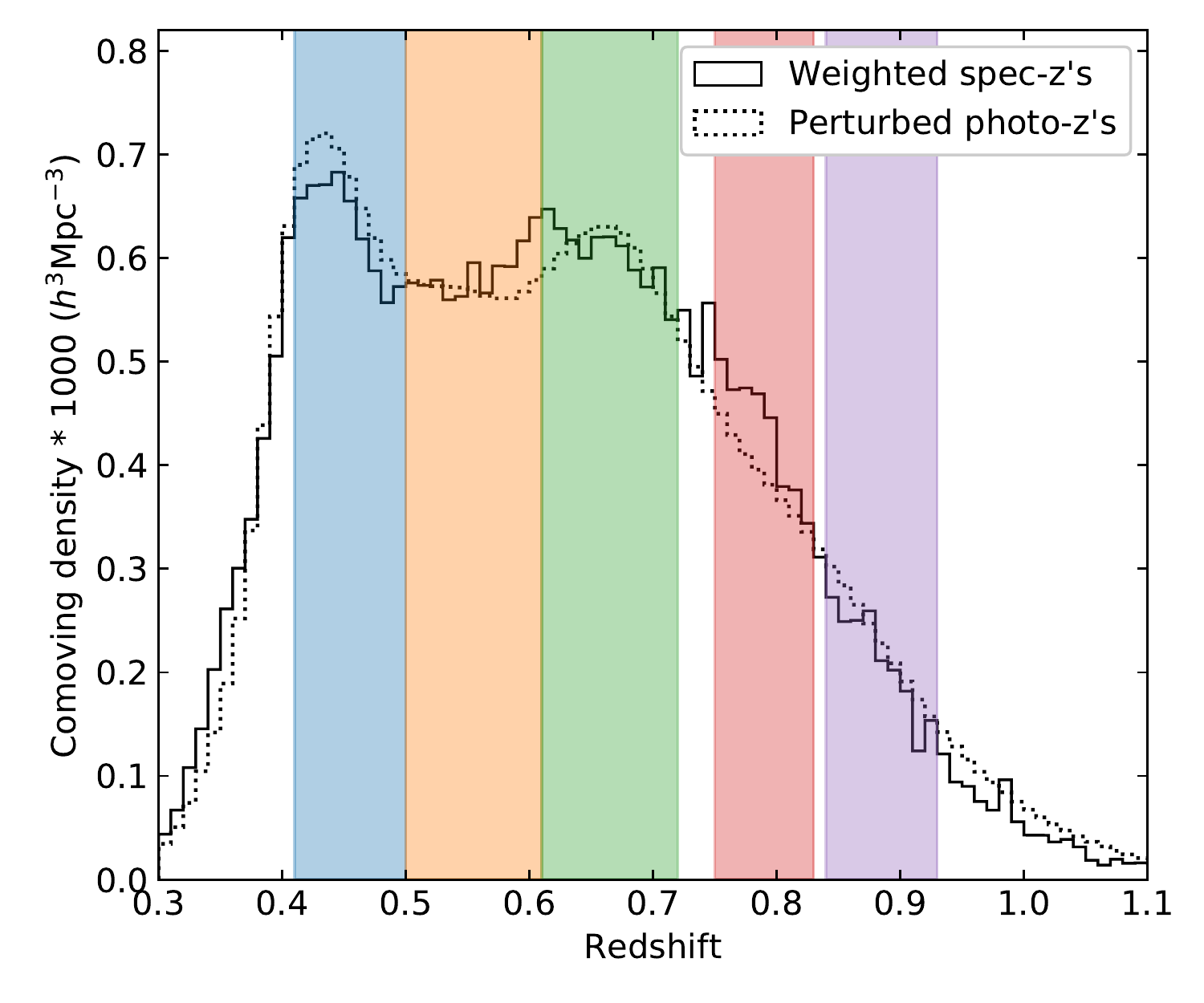}
    \centering
    \caption{Redshift-dependent comoving number density of the LRG sample. The dashed line shows the densities estimated with photo-$z$'s convolved by the estimated photo-$z$ error. The solid line shows the densities estimated from the weighted spec-$z$'s described in \ref{sec:validation}; these weighted spec-$z$' are also used for deriving the comoving number densities for the clustering analysis. The colored bands represent the redshift bins. The fact that the two lines are consistent with each other indicates that our comoving density estimates are relatively robust. The uncertainties in the densities are accounted for in our modeling (see section \ref{sec:mcmc_sampling})}
    \label{fig:lrg_comoving_density}
\end{figure}

In order to study the redshift dependence of the sample properties, we divide the LRG sample into four photometric redshift bins of width $\Delta z \simeq 0.1$, covering the ranges [0.41, 0.5], [0.5, 0.61], [0.61, 0.72], [0.75, 0.83], [0.84, 0.93]. These bins have been chosen such that they are centered at the redshifts of the snapshots of the halo catalogs from the N-body simulation which we use for the clustering analysis in  section \ref{sec:modeling} (there are small gaps between some of the redshift bins for the same reason). Fig. \ref{fig:lrg_comoving_density} shows the estimated comoving number density vs. redshift for the LRG sample; the shaded regions represent the  redshift ranges for each bin. The volume-averaged comoving number densities for each bin are listed in Table \ref{tab:redshift_bins}. The densities estimated from photo-$z$'s are consistent with the estimates from weighted spec-$z$'s. Densities estimated using spec-$z$'s weighted as described in section \ref{sec:validation} are used in modeling.

\begin{table}
    \centering
    \caption{The redshift bins. The second column lists the redshifts of the snapshots of the N-body simulation. The third column lists the comoving number densities in units of $h^{3}\mathrm{Mpc}^{-3}$; these values are used in the HOD fitting with $10\%$ assumed Gaussian uncertainty.}
    \label{tab:redshift_bins}
    \begin{tabular}{ccc}
    \hline
    Redshift & $z_{\mathrm{sim}}$ & $n(z)$\\ \hline
    $0.41<z_{\mathrm{phot}}<0.50$ & 0.4573 & $6.32 \times 10^{-4}$ \\
    $0.50<z_{\mathrm{phot}}<0.61$ & 0.5574 & $6.16 \times 10^{-4}$ \\
    $0.61<z_{\mathrm{phot}}<0.72$ & 0.6644 & $6.15 \times 10^{-4}$ \\
    $0.75<z_{\mathrm{phot}}<0.83$ & 0.7787 & $4.41 \times 10^{-4}$ \\
    $0.84<z_{\mathrm{phot}}<0.93$ & 0.8594 & $2.14 \times 10^{-4}$ \\
    \hline
    \end{tabular}
\end{table}

Fig. \ref{fig:lrg_surface_density} shows the redshift distribution (surface density) for each redshift bin estimated by convolving the photo-$z$'s with their photo-$z$ error.

\subsection{Projected correlation function}
\label{sec:wprp}
Ideally, one would like to measure the correlation function $\xi(r)$, the excess probability of finding a pair of galaxies separated by distance $r$, but the large uncertainty in radial distances precludes its direct measurement. For imaging datasets, it is common to measure instead the angular correlation function, typically in bins of photometric redshifts. However, in doing so the information on relative distances between galaxies contained in the photo-$z$'s is not utilized. For spectroscopic datasets, due to the presence of redshift-space distortions, the small-scale clustering is usually measured with the projected correction function, effectively integrating out the effects of redshift-space distortions:
\begin{equation}
\label{eq:wprp}
w_\mathrm{p}(r_\mathrm{p}) = \int_{-\pi_\mathrm{max}}^{\pi_\mathrm{max}} \xi(r_\mathrm{p}, \pi) d\pi,
\end{equation}
where $\xi(r_\mathrm{p}, \pi)$ is the 3-D correlation function, $r_\mathrm{p}$ is the transverse distance and $\pi$ is the line-of-sight distance.

\begin{figure}
    \centering
    \includegraphics[width=1.02\columnwidth]{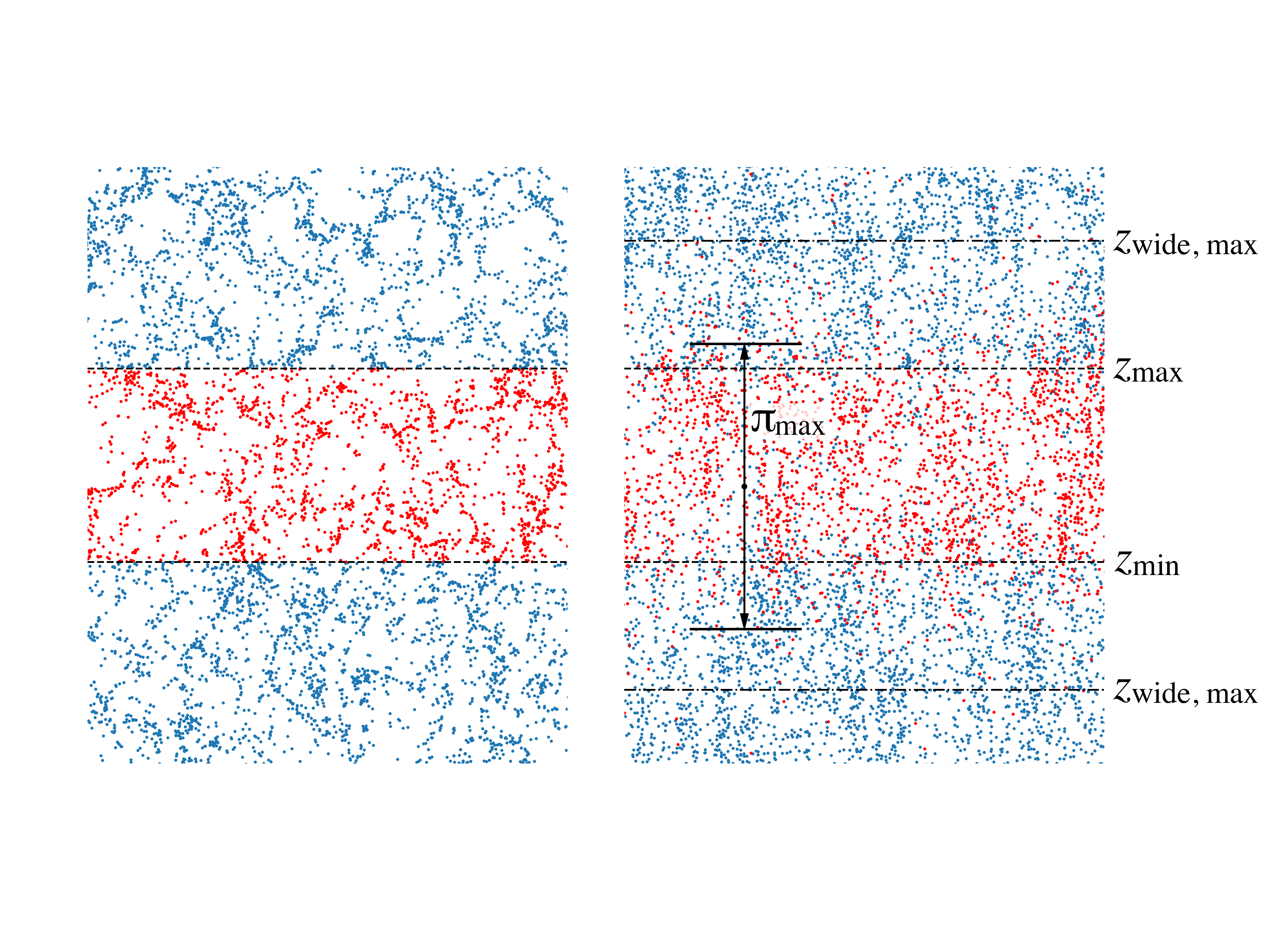}
    \caption{Illustration of our method for measuring $w_\mathrm{p}(r_\mathrm{p})$ with photo-$z$'s; differences in the redshift direction are shown to scale for the $0.61 < z_\mathrm{phot} < 0.72$ redshift bin. \textit{Left panel}: galaxies with accurate redshifts. The dashed lines show the redshift bin boundary, and the galaxies inside and outside the redshift bin are color-coded in red and blue, respectively. \textit{Right panel}: the same galaxies with the same color coding as the left panel, but with realistic redshift (photo-$z$) uncertainties added. A significant number of galaxies cross the redshift boundaries. The arrows show the $\pm \pi_\mathrm{max}$ range with $\pi_\mathrm{max}=150 h^{-1}\mathrm{Mpc}$ used in counting pairs and calculating $w_\mathrm{p}(r_\mathrm{p})$. The dash-dotted line shows the wider redshift bin used in the ``padded'' $w_\mathrm{p}(r_\mathrm{p})$ calculation; by construction it includes all objects that are within $\pm \pi_\mathrm{max}$ of any object with $0.61 < z_\mathrm{phot} < 0.72$.}
    \label{fig:wprp_illustration}
\end{figure}

Here, to better exploit the photo-$z$ information, we measure the projected correction function (Equation \ref{eq:wprp}), using distances derived from the photo-$z$'s. Fig. \ref{fig:wprp_illustration} illustrates our method of measuring $w_\mathrm{p}(r_\mathrm{p})$. We adopt a relatively large $\pi_\mathrm{max}$ of $150 h^{-1}\mathrm{Mpc}$ to account for the large radial distance uncertainties from the photo-$z$'s. This large $\pi_\mathrm{max}$ is comparable to the width of the redshift bins ($160 - 210 h^{-1}\mathrm{Mpc}$). However, the photo-$z$ errors cause many galaxy pairs to be lost due to one of the galaxies being outside of the redshift bin in a generic auto-correlation measurement, thus complicating the modeling and resulting in a lower clustering signal-to-noise. To address this issue, instead of counting pairs within the redshift bin $z_i$ (i.e., measuring a simple auto-correlation), we define a wider redshift bin $z_{\mathrm{wide}, i}$ that encloses $z_i$, and count pairs between galaxies in $z_i$ and galaxies in $z_{\mathrm{wide}, i}$. The wider redshift bin $z_{\mathrm{wide}, i}$ extends from $z_i$ by $\pi_\mathrm{max}$ in both directions, so that for all galaxies in $z_{\mathrm{wide}, i}$, all their pairs within $\pi_\mathrm{max}$ will be counted. Besides boosting the clustering S/N, this ``padded'' auto-correlation approach decouples the clustering measurement from any effects associated with the boundary of the redshift bin, and thus simplifies the measurement and modeling of the projected correlation function.

To compute this ``padded'' $w_\mathrm{p}(r_\mathrm{p})$, we adopt the cross-correlation form of the Landy-Szalay estimator \citep{landy_bias_1993}:
\begin{equation}
\label{eq:nolabel11}
w_\mathrm{p}(r_\mathrm{p}) = 2 \pi_\mathrm{max} \times \frac{(D_1 D_2 - D_1 R_2 - D_2 R_1 + R_1 R_2)}{R_1 R_2},
\end{equation}
where each term denotes the pair count between two samples; $D_1$ denotes the data (galaxy sample) in a redshift bin defined in section \ref{sec:redshift_bins} and $D_2$ denotes the data in the wider redshift bin; and $R_1$ and $R_2$ denote sets of random points with the same angular and redshift distribution as $D_1$ and $D_2$, respectively. The redshifts of the randoms are randomly drawn from the redshifts of the data, so they have the same redshift distribution as the data, by construction. In each redshift bin the ratio of the number of randoms to data is 20. The measurement is done using the \textsc{TreeCorr} package \citep{jarvis_skewness_2004}. We tested this estimator on mocks that resemble our redshift bins, and we confirmed that it produces the same $w_\mathrm{p}(r_\mathrm{p})$ as that for mocks in a cubic volume with periodic boundary conditions using the simple auto-correlation estimator.

\begin{figure*}
    \includegraphics[width=0.99\textwidth]{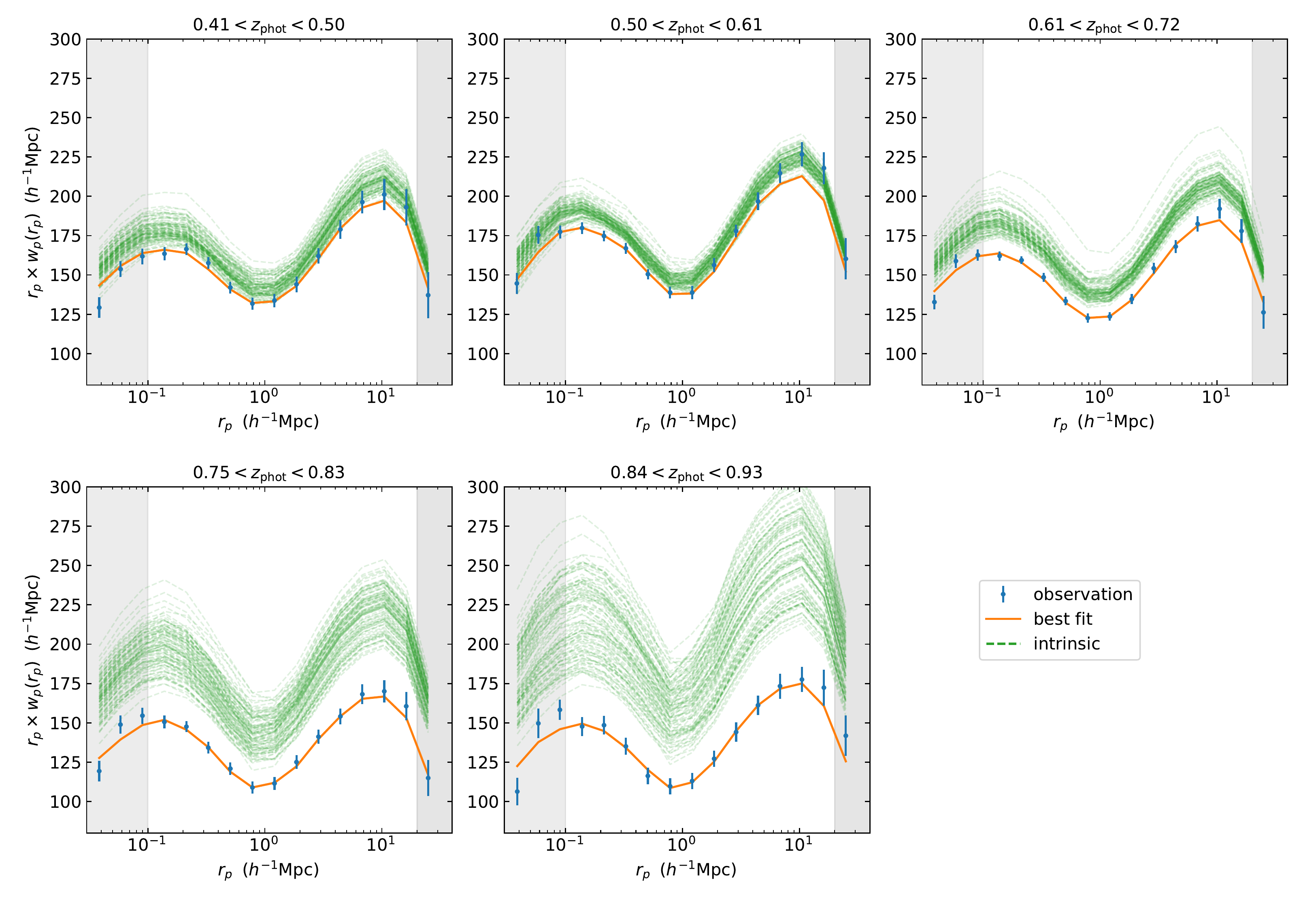}
    \centering
    \caption{The projected correlation function multiplied by the transverse distance. The points are measurements with error bars from jackknife. The points in the gray shaded area are not used for modeling to avoid possible systematics. The orange curve is the best fit from HOD modeling. The green band is the [16th, 84th] percentile range of the intrinsic clustering signal, i.e., the clustering that would have been measured according to the fit parameters if perfect distance measurements (rather than photo-$z$'s) were available. The larger uncertainty in the intrinsic clustering in the higher redshift bins is mostly due to the larger photo-$z$ errors for them.}
    \label{fig:wprp}
\end{figure*}

We measure $w_\mathrm{p}(r_\mathrm{p})$ in 12 logarithmically-spaced bins covering the range from $0.11 h^{-1}\mathrm{Mpc}$ to $19.5 h^{-1}\mathrm{Mpc}$ (comoving). We also made measurements at smaller and larger scales, but these measurements were not used for the modeling. Fig. \ref{fig:wprp} shows the measured projected correlation functions for the five redshift bins. The two bumps, corresponding to the one-halo (from galaxy pairs within the same dark matter halo) and two-halo (from galaxy pairs in two different halos) regimes of the correlation function, are clearly visible in all the redshift bins. It should be noted that the photo-$z$ errors effectively smooth out sharp features in the radial direction, and as a result these $w_\mathrm{p}(r_\mathrm{p})$ measurements using photo-$z$'s have lower amplitude than the intrinsic clustering signal that one can measure with spectroscopic redshifts.

\begin{figure}
    \includegraphics[width=0.95\columnwidth]{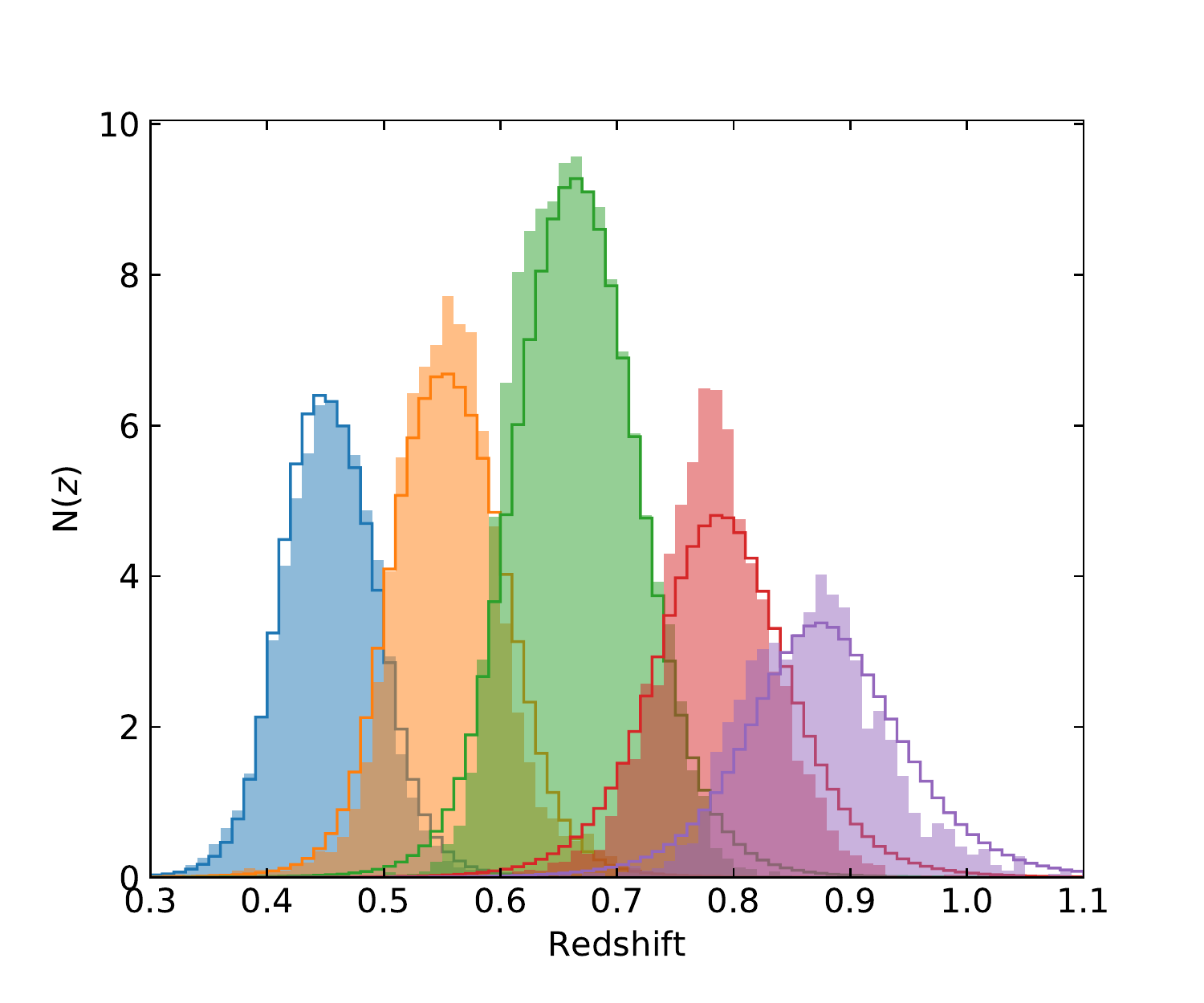}
    \centering
    \caption{The estimated redshift distributions (surface density in redshift bins) of the LRG sample. The y-axis is the number of objects per sq. degree in the redshift bin of width 0.1. The filled histograms show the redshift distributions estimated from weighted spec-$z$ objects. The unfilled histograms are distributions of stacked photo-$z$'s that are convolved with photo-$z$ errors (however, objects are assigned to a given redshift bin based only upon their random forest point estimates). The samples in different redshift (photo-$z$) bins overlap due to photo-$z$ errors. Note that 1) we use the original photo-$z$ error estimates, which as seen in Fig. \ref{fig:pz_error_validation}, are probably overestimated; and 2) this simple stacking of the photo-$z$ probabilities produces broader redshift distributions than the mathematically correct procedure (see \citealt{malz_how_2020}).}
    \label{fig:lrg_surface_density}
\end{figure}

As shown in Fig. \ref{fig:lrg_surface_density}, the sample in each redshift bin does not contain all the objects in that redshift range, and it also includes objects whose true redshift is in other bins.

\subsection{Jackknife resampling and covariances}
\label{sec:jackknife}
We compute the covariance matrices of correlation functions with jackknife resampling: we divide the footprint into $N_\mathrm{{sub}}$ subregions, and we resample the dataset with one of the subregions removed. There are a total of $N_\mathrm{{sub}}$ resampled datasets, and the correlation function is measured for each one. The covariance matrix is given by
\begin{equation}
\label{eq:nolabel12}
    \mathrm{Cov}(w_{i}, w_{j})=\frac{(N_\mathrm{sub}-1)}{N_\mathrm{sub}} \sum_{l=1}^{N_\mathrm{{sub}}}\left(w_{i}^{l}-\overline{w}_{i}\right)\left(w_{j}^{l}-\overline{w}_{j}\right),
\end{equation}
where $w_{i}^{l}$ is the projected correlation function at the $i$-th distance bin measured from the $l$-th jackknife sample, and $\overline{w}_{i}$ is the mean from all jackknife samples.

The DECaLS survey was not yet completed by DR7. This, combined with the masks and quality cuts, results in the irregular survey geometry, making it difficult to manually divide the footprint into compact subregions with equal areas. To solve this problem, we developed an automated routine. The relevant Python codes can be found online
\footnote{\url{https://github.com/rongpu/pixel_partition}}
. The routine involves three steps:
\begin{enumerate}
    \item The objects (in our case randoms) in the survey footprint are divided into HEALPix pixels. This significantly reduces the number of points and speeds up computation;
    \item Initial grouping of the HEALPix pixels is performed using a clustering algorithm (specifically, we use k-means clustering; see \citealt{hartigan_algorithm_1979}) with the object count in each pixel as weights to account for fractional occupation of the pixels;
    \item We change the group labels of a specific fraction of boundary pixels, which were of randomly selected, to the labels of their neighboring group(s), and only keep these changes if they improve the ``score'' which is a weighted sum of the compactness (defined as the average of the standard deviations of the angular distances, in arcseconds, between the positions of all objects in a subregion and its center) and uniformity of the areas (defined as the standard deviation of the areas of the subregions);
    \item We repeat this step until the desired score is achieved.
\end{enumerate}
Specifically, in each iteration the fraction of changed boundary pixels is 0.15\%; and equal weights were given to uniformity and compactness for calculating the score.

\begin{figure*}
    \includegraphics[width=0.99\textwidth]{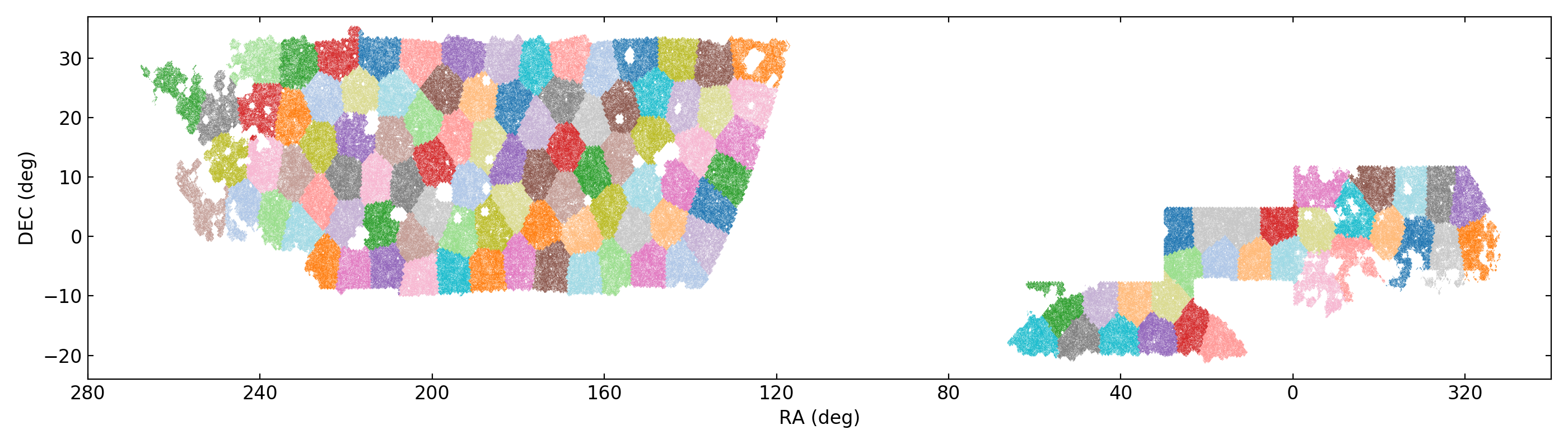}
    \caption{Sky distribution of the LRGs with color coding to show the 120 subregions for jackknife resampling. Note that nonadjacent ``patches'' with the same color are different subregions. Each subregion has the same area (within ${\sim}\,2\%$ variation) and is compact by design.}
    \label{fig:sky_jackknife}
\end{figure*}

Applying this procedure, we divide the footprint into 120 jackknife subregions of $47.1$ sq. degrees each (within ${\sim}\,2\%$ variation), shown in Fig. \ref{fig:sky_jackknife}. The areas of the subregions are uniform to ${\sim}\,1\%$. The area of the subregions is large enough to cover the entirety of angular scales of interest, and small enough to produce enough subregions to compute covariance matrices accurately. The size of each subregion is much larger than the largest angular scale of ${\sim}\,1$ degree in the correlation function, therefore the jackknife resampling accounts for cosmic variance in addition to shot noise and imaging systematics.

\section{Modeling}
\label{sec:modeling}

In this section we present our analysis of the galaxy-halo connection for the LRG sample using the HOD framework. The galaxy-halo connection determines the clustering properties of the galaxy sample, such as the large-scale bias, which is useful for forecasting DESI BAO constraints. Predictions of the locations of galaxies from simulations of dark matter alone also forms the basis of mock galaxy catalogs that can be used for estimating covariances between measured large-scale-structure statistics. In this work we fix the cosmological parameters and only allow the HOD parameters to vary, but in principle the method presented here can be incorporated into a more flexible modeling framework that constrains both cosmology and galaxy-halo connection parameters.

\subsection{HOD model}
\label{sec:hod}
We fit the measured clustering signal with an HOD model (e.g., see \citealt{berlind_halo_2002} which also lists earlier literature on HOD; see also \citealt{wechsler_connection_2018} for a more general review on the galaxy-halo connection), which is widely used to model luminosity-threshold galaxy samples. In this framework, dark matter halos from N-body simulations are populated by central and satellite galaxies with a probabilistic prescription. In its basic form which we have adopted, the central galaxy probability (denoted by $N_{\mathrm{cen}}$) and the mean number of satellite galaxies (denoted by $N_{\mathrm{sat}}$) for a given a halo are determined solely by the halo mass. There are several slightly different mathematical prescriptions for this. We adopt the one in \citet{zentner_constraints_2019}; we briefly summarize it below.

In this formulation, the central galaxy probability is given by a step-like function
\begin{equation}
\label{eq:nolabel13}
    \left\langle N_{\mathrm{cen}} | M_{\mathrm{vir}}\right\rangle=\frac{1}{2}\left(1+\operatorname{erf}\left[\frac{\log \left(M_{\mathrm{vir}}\right)-\log \left(M_{\min}\right)}{\sigma_{\log M}}\right]\right),
\end{equation}
where $M_{\mathrm{vir}}$ is the virial mass of the dark matter halo; $M_{\min}$ is the mass threshold above which halos are populated by central galaxies; and $\sigma_{\log M}$ defines the smoothness of this transition;

The number of satellite galaxies in a halo follows the Poisson distribution with a mean given by a power law
\begin{equation}
\label{eq:nolabel14}
    \left\langle N_{\mathrm{sat}} | M_{\mathrm{vir}}\right\rangle=\left(\frac{M_{\mathrm{vir}}-M_0}{M_1}\right)^{\alpha},
\end{equation}
where $M_0$, $M_1$ and $\alpha$ are free parameters of the HOD model. We impose that there are no satellite galaxies in halos with $M_{\mathrm{vir}}<M_0$. We allow satellite galaxies to exist in halos with no central galaxies.

The spatial distribution of satellite galaxies is assumed to follow the NFW profile \citep{navarro_universal_1997}, with the scale radius $R_s$ given by the value for the parent halo in the N-body catalog. We ignore the effect of velocity dispersion on redshift since it is negligible compared to photo-$z$ errors.

\subsection{Mock galaxies}
\label{sec:mock}

To constrain the HOD model parameters, we measure the clustering of mock galaxies generated from a set of assumed values for these parameters and compare with our measurements from real LRGs. We use the halo catalog from the MultiDark Planck 2 (MDPL2) simulation \citep{klypin_multidark_2016}, which used Rockstar \citep{behroozi_rockstar_2013} to identify the halos. The MDPL2 simulation adopts the Planck 2013 cosmology \citep{planckcollaboration_planck_2014}: $\Omega_\mathrm{m}=0.307115$, $\Omega_\Lambda=1-\Omega_\mathrm{m}=0.692885$, $\Omega_b=0.048206$, $h=0.6777$, $\sigma_8=0.823$, and $\mathrm{n_s}=0.96$. The size of the cubic simulation box is $1 h^{-1}\mathrm{Gpc}$, and the mass resolution is $1.51\times10^9 h^{-1}M_\odot$. We designed the redshift bins of the LRG sample so that they center at the redshifts of five snapshots of the simulation, as listed in Table \ref{tab:redshift_bins}. We use \textsc{halotools} \citep{hearin_forward_2017} for populating the halos with galaxies using the pre-defined \citet{zheng_galaxy_2007} prescription (note that we use the default \textsc{halotools} definition of $\left\langle N_{\mathrm{sat}} | M_{\mathrm{vir}}\right\rangle$ which is is slightly different from \citealt{zheng_galaxy_2007}).

To emulate the effect of photo-$z$'s on the clustering signal, we perturb the position of the galaxies along the line-of-sight direction, i.e., the direction along one of the axes of the simulation box. For each galaxy, this distance perturbation is drawn from a Gaussian distribution, the width of which is randomly drawn from the rescaled photo-$z$ error estimates of the LRGs in the corresponding redshift bin. As discussed in section \ref{sec:pz_error_validation}, the true photo-$z$ errors are assumed to differ from the estimated errors by a scaling factor $S_z$ which we do not have good constraints on. Thus $S_z$ is included in the model as a nuisance parameter. We use the \textsc{Corrfunc} software package \citep{2017ascl.soft03003S} for measuring $w_\mathrm{p}(r_\mathrm{p})$ of the mock galaxies because it is better optimized than \textsc{TreeCorr} for cubic boxes with periodic boundaries conditions.

To test whether allowing skewness in the photo-$z$ PDF would change our results, we have repeated the analysis using a skew normal distribution as the photo-$z$ PDF for one of the redshift bins, with skewness derived by fitting the $(z_{\mathrm{phot}}-z_{\mathrm{spec}})/\sigma_z$ distribution. We find negligible changes in the results; see Appendix \ref{sec:skewnorm} for more details.

\subsection{MCMC sampling of parameters}
\label{sec:mcmc_sampling}

As described in section \ref{sec:hod}, the HOD model used here has five free parameters: $\mathrm{log}(M_{\min})$, $\sigma_{\log M}$, $\alpha$, $\mathrm{log}(M_0)$, and $\mathrm{log}(M_1)$. Additionally, we have the nuisance parameter $S_z$ to account for uncertainties in the photo-$z$ error estimation. We adopt flat priors for all these parameters, and the ranges for the priors are listed in Table \ref{tab:priors}. We set the lower limit of $S_z$ at 0.6, which is much lower than the values inferred from spectroscopic redshifts (see Fig. \ref{fig:pz_error_validation}).

\begin{table}
    \centering
    \caption{The ranges of the flat priors on model parameters.}
    \label{tab:priors}
    \begin{tabular}{ll}
        \hline
        Parameter & Prior Interval\\
        \hline
        $\mathrm{log}(M_{\min})$   &   [11.0, 14.0] \\
        $\sigma_{\log M}$   &    [0.001, 1.5] \\
        $\alpha$   &    [0.0, 2.0] \\
        $\mathrm{log}(M_0)$   &    [11.0, 14.0] \\
        $\mathrm{log}(M_1)$   &   [11.5, 15.5] \\
        $S_z$   &    [0.6, 1.4] \\        \hline
    \end{tabular}
\end{table}

To obtain the posterior probability distributions of the HOD parameters, we perform Markov Chain Monte Carlo (MCMC) sampling using the \textbf{emcee} package \citep{foreman-mackey_emcee_2013}. The likelihood function used is given by $\mathcal{L} \propto e^{-\chi^{2}/2}$, where $\chi^{2}$ is given by
\begin{equation}
\label{eq:nolabel15}
    \chi^{2}=\Delta w_{i}\left[\mathrm{Cov}^{-1}\right]_{i j} \Delta w_{j}+\frac{\left(n^{\mathrm{meas}}-n^{\mathrm{mock}}\right)^{2}}{\sigma_n^{2}},
\end{equation}
where $\Delta w_{i}=w_p^{meas}(r_\mathrm{p})-w_p^{mock}(r_\mathrm{p})$ is the difference between the measured projected correlation function and the one from the mocks at the $i$-th distance bin; $\mathrm{Cov}^{-1}$ is the inverse of the covariance matrix from jackknife resampling (see section \ref{sec:jackknife}); $n^{\mathrm{meas}}$ and $n^{\mathrm{mock}}$ are the comoving number densities of  the data and the mock, respectively; and $\sigma_n$ is the uncertainty in the comoving number densities. The comoving number densities are estimated from the weighted spec-$z$ objects as described in \ref{sec:validation}. We have compared these density estimates with the densities estimated from the photo-$z$'s, and the standard deviation of the differences in the five redshift bins is $\sim 7\%$. We adopt a larger density uncertainty of $\sigma_n=10\%$ to account for any additional unknown systematics in the photo-$z$'s.

\section{Results}
\label{sec:mcmc_results}

\begin{figure*}
   \includegraphics[width=0.97\textwidth]{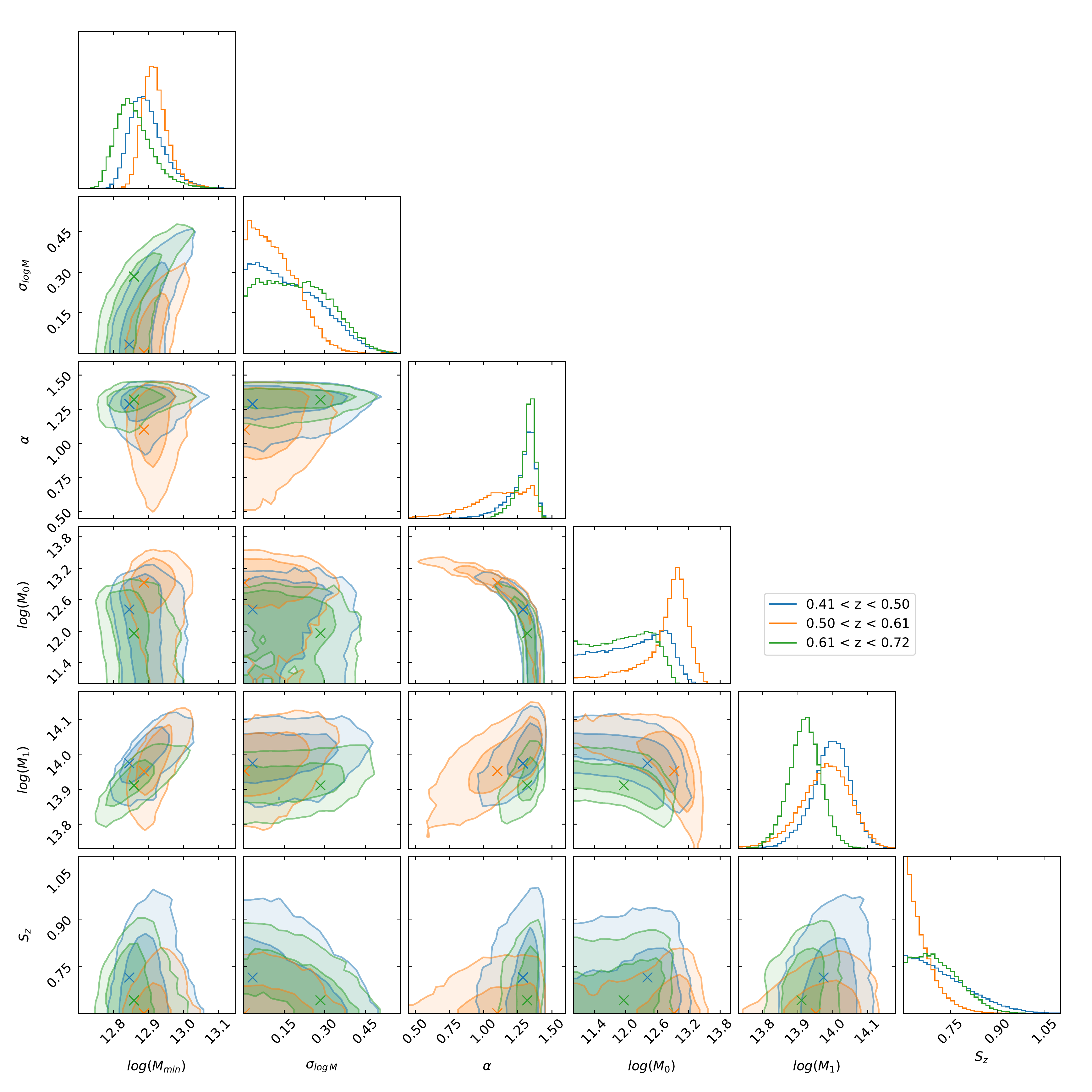}
   \centering
    \caption{The one- and two-dimensional marginalized posterior probability distributions of the model parameters from MCMC for the first three redshift bins. The inner contour and outer contour are the 68\% and 95\% confidence regions, respectively. All parameters except $S_z$ are HOD parameters; the parameter $S_z$ is the scaling factor for the photo-$z$ errors. The mass parameters are in units of $h^{-1}M_\odot$.}
    \label{fig:mcmc_1}
\end{figure*}

\begin{figure*}
   \includegraphics[width=0.97\textwidth]{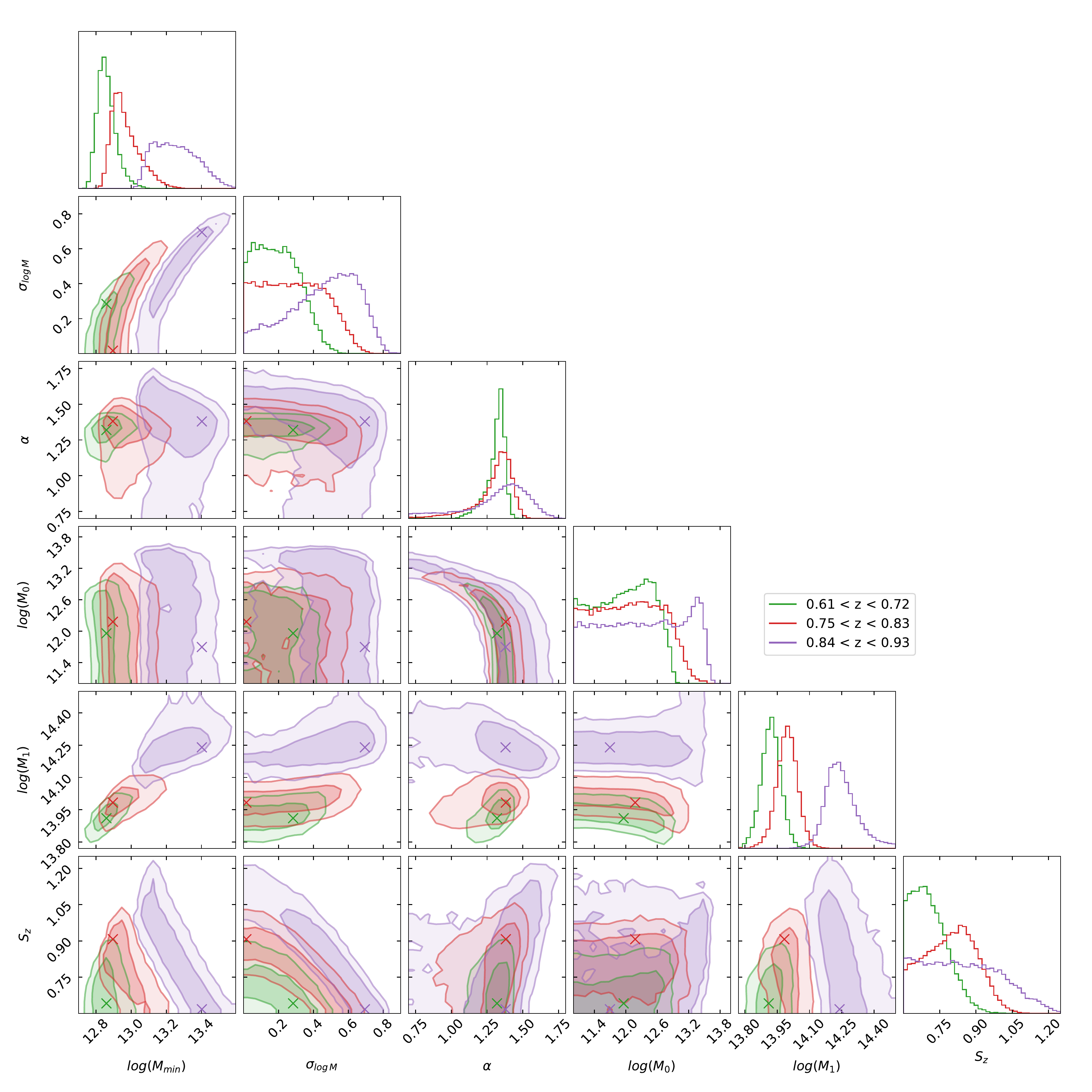}
   \centering
    \caption{Same as Fig. \ref{fig:mcmc_1}, but for different redshift bins. We include the results of $0.61 < z_\mathrm{phot} < 0.72$ (which was also shown in the previous plot). The contours of the highest-redshift bin ($0.84 < z_\mathrm{phot} < 0.93$) are significantly different from those of the other redshift bins.}
    \label{fig:mcmc_2}
\end{figure*}

The posterior distributions of the model parameters are shown in Fig. \ref{fig:mcmc_1} and Fig. \ref{fig:mcmc_2}. These plots have been made using a modified version of the \textsc{corner.py} software package \citep{foreman-mackey_corner.py_2016}. The mean, median and best-fit values of the parameters and the 16\% and 84\% percentiles are listed in Table \ref{tab:posterior}. The derivation of the best-fit parameters is described in Appendix \ref{sec:noisy_mcmc}. The first four redshift bins have very similar HOD parameters, indicating that the LRGs have similar host halo properties over the redshift range of $0.4<z<0.8$.

The highest-redshift sample has much wider contours on the model parameters than lower redshift samples. This could be due to a combination of several factors, including the larger photo-$z$ errors at higher redshifts, a smaller sample size, and a broader redshift distribution that dilutes the clustering signal. Despite the larger errors, the differences in the model parameters between this and lower redshift samples are statistically significant. Such deviation is caused by the selection effect of the apparent magnitude limit (see the middle panel of Fig. \ref{fig:sample_selection}): at redshifts higher than ${\sim}\,0.7$, the luminosity-threshold established by the sliding cut is replaced by the apparent magnitude limit, and as a result more luminous galaxies are more and more preferentially selected as redshift increases. This qualitative change of the sample at high redshift will need to be considered when analyzing and interpreting DESI spectroscopic data.

At lower redshifts where the photo-$z$ errors are small, the HOD parameters are not sensitive to the photo-$z$ error rescaling factor $S_z$, since most of the galaxy pairs are still within $\pi_\mathrm{max} = 150 h^{-1}\mathrm{Mpc}$ in the line-of-sight direction. At higher redshifts the photo-$z$ errors are much larger, and the clustering signal from the mock galaxies is much more sensitive to the $S_z$, thus resulting in the strong correlation between some of the HOD parameters and $S_z$. Nevertheless, the value of $S_z$ is poorly constrained by the data, although the posterior prefers $S_z \lesssim 1$ at all redshifts, which is consistent with the results of the photo-$z$ error validation with spectroscopic redshifts.

\begin{figure*}
    \includegraphics[width=0.97\textwidth]{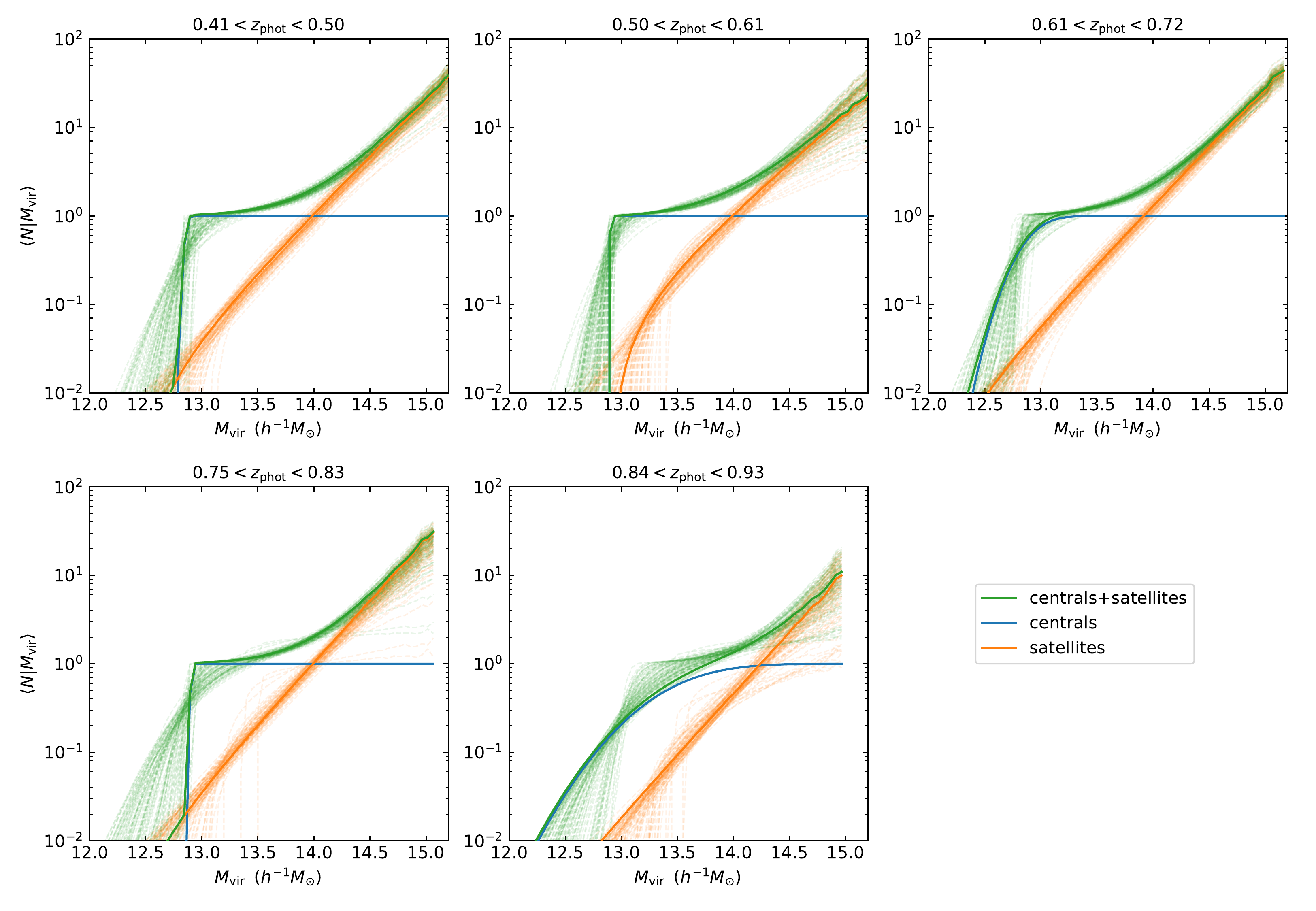}
    \centering
    \caption{The halo occupations, i.e., the average number of galaxies that are hosted by a halo of a certain mass. The occupations of centrals, satellites and the full occupation are plotted separately. The solid lines are from the best-fit parameters. The dashed lines are from 100 sets of parameters randomly selected from the MCMC chain (only the satellite and full occupations are plotted for clarity).}
    \label{fig:hod_occupation}
\end{figure*}

\begin{figure*}
    \includegraphics[width=0.97\textwidth]{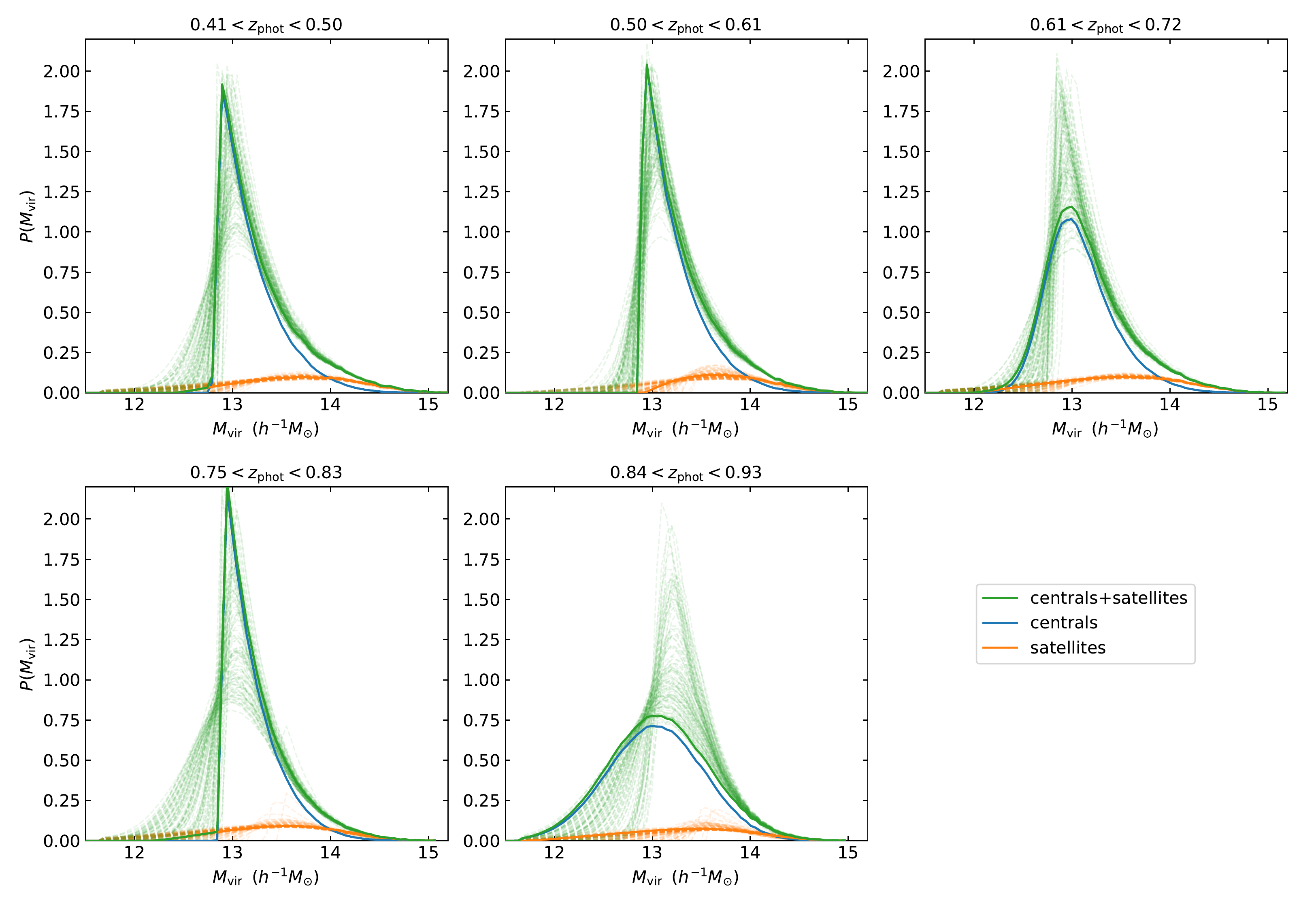}
    \centering
    \caption{Similar to Fig. \ref{fig:hod_occupation}, but showing the probability distribution of host halo mass for a randomly selected galaxy in each redshift bin, with the solid lines showing the best-fit results and dashed lines showing parameters from the MCMC chain. The green curve shows the host halo mass distribution for all galaxies (centrals+satellites) normalized as the probability per $\mathrm{log}M$; the blue and orange curves show the central and satellite components. The centrals in the highest-redshift sample have a much smoother low-mass cut-off than in the lower redshift samples.}
    \label{fig:hod_distribution}
\end{figure*}

Fig. \ref{fig:hod_occupation} shows the halo occupation functions. Fig. \ref{fig:hod_distribution} shows the probability distributions for the host halo masses in each redshift bin. In both figures, the solid lines are the best-fit results, and the dashed lines are for parameters randomly selected from the MCMC chains to show the possible range of halo occupations allowed by the data. The particularly sharp transition (quantified by the $\sigma_{\log M}$ parameter) in the central galaxy population might be due to insufficient flexibility in the HOD model or inaccuracy in the $\sigma_8$ parameter we assumed, and the strong correlation between $\sigma_{\log M}$ and the photo-$z$ error rescaling factor also suggests the possibility of biased $\sigma_{\log M}$ estimates due to additional photo-$z$ systematics that is not accounted for by the rescaling factor.

\begin{figure*}
    \includegraphics[width=0.99\textwidth]{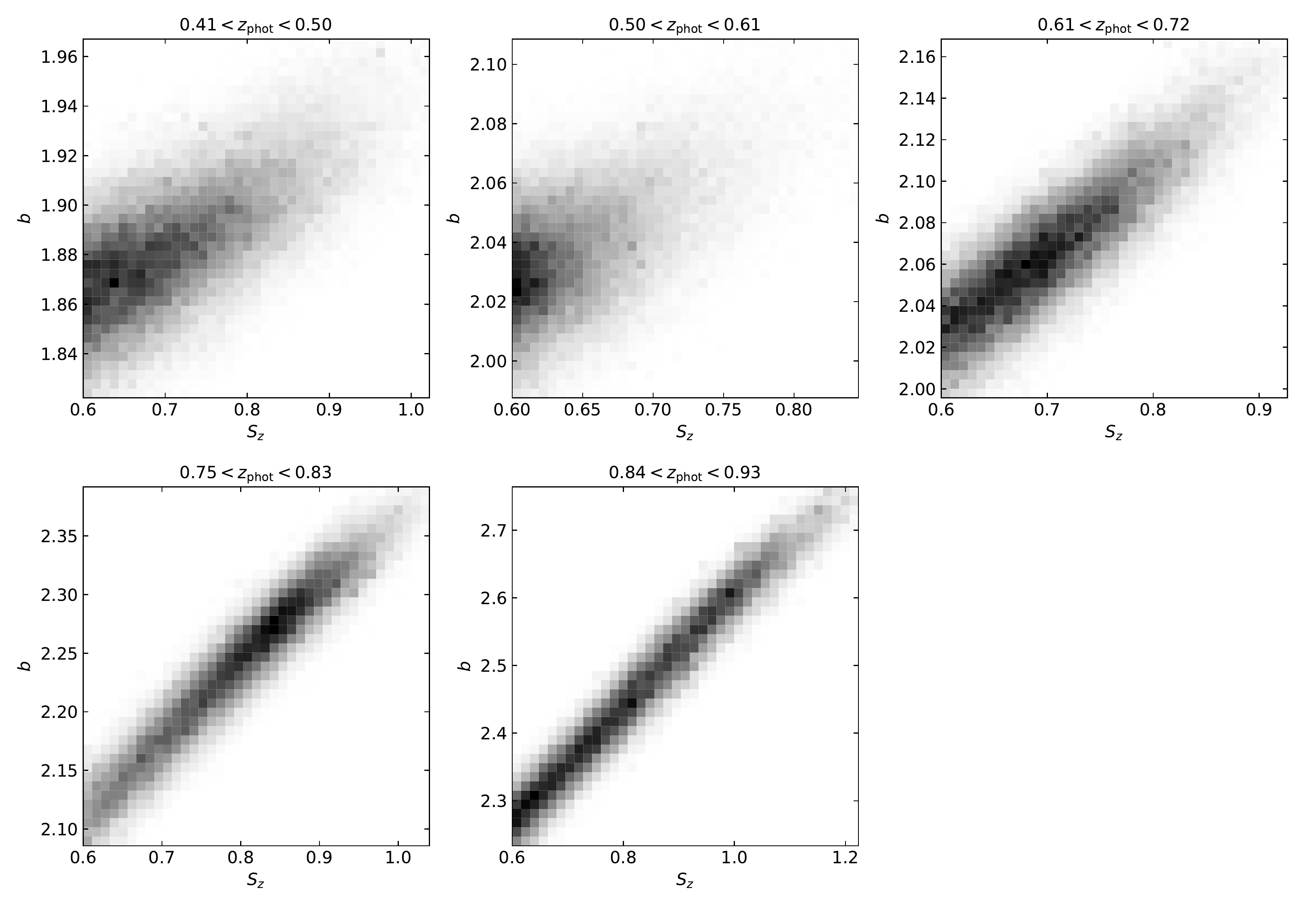}
    \centering
    \caption{Two dimensional posterior distribution of galaxy bias vs. the photo-$z$ error rescaling factor $S_z$ (a nuisance parameter in our analysis) at different redshifts. The correlations are much stronger at higher redshifts due to the overall larger photo-$z$ errors that dilute the clustering signal.}
    \label{fig:bias_sz_correlation}
\end{figure*}

Ideally, we would compute the galaxy bias by comparing the clustering amplitudes of galaxies and dark matter particles in the N-body simulation as below
\begin{equation}
\label{eq:nolabel16}
    b = \left(\xi_\mathrm{gal}/\xi_\mathrm{matter}\right)^{1/2},
\end{equation}
where $\xi_\mathrm{gal}$ and $\xi_\mathrm{matter}$ are the two-point correlation functions of galaxies and matter, respectively. (The galaxy bias is a function of scale, and at large scale it asymptotes to the large-scale bias value which we compute here; hereafter we refer to the large-scale bias simply as the galaxy bias and denote it as $b$.) However, due to the lack of access to the MDPL2 dark matter particle catalog, we instead use the analytic halo bias-halo mass relation from \citet{tinker_large_2010}, implemented by the \textsc{Colossus} software package \citep{diemer_colossus_2018}. Specifically, we obtain the galaxy bias by averaging over the bias of all halos weighted by the number of galaxies in each halo,
\begin{equation}
    b_\mathrm{gal} = \frac{1}{N_\mathrm{gal}}\sum_{i} b_\mathrm{halo}(M_\mathrm{vir, i}) N_\mathrm{gal}(i)
\end{equation}
where $N_\mathrm{gal}$ is the total number of galaxies, $M_\mathrm{vir, i}$ is the virial mass of the $i$-th halo, and $N_\mathrm{gal}(i)$ is the number of galaxies (central and satellites) in the $i$-th halo.

There is a strong degeneracy between $S_z$ and galaxy bias, as shown in Fig. \ref{fig:bias_sz_correlation}, which leads to large uncertainties in galaxy bias at high redshift. This degeneracy is expected, since a stronger clustering signal could be due to either a higher bias or smaller photo-$z$ errors.

\begin{figure}
    \includegraphics[width=0.95\columnwidth]{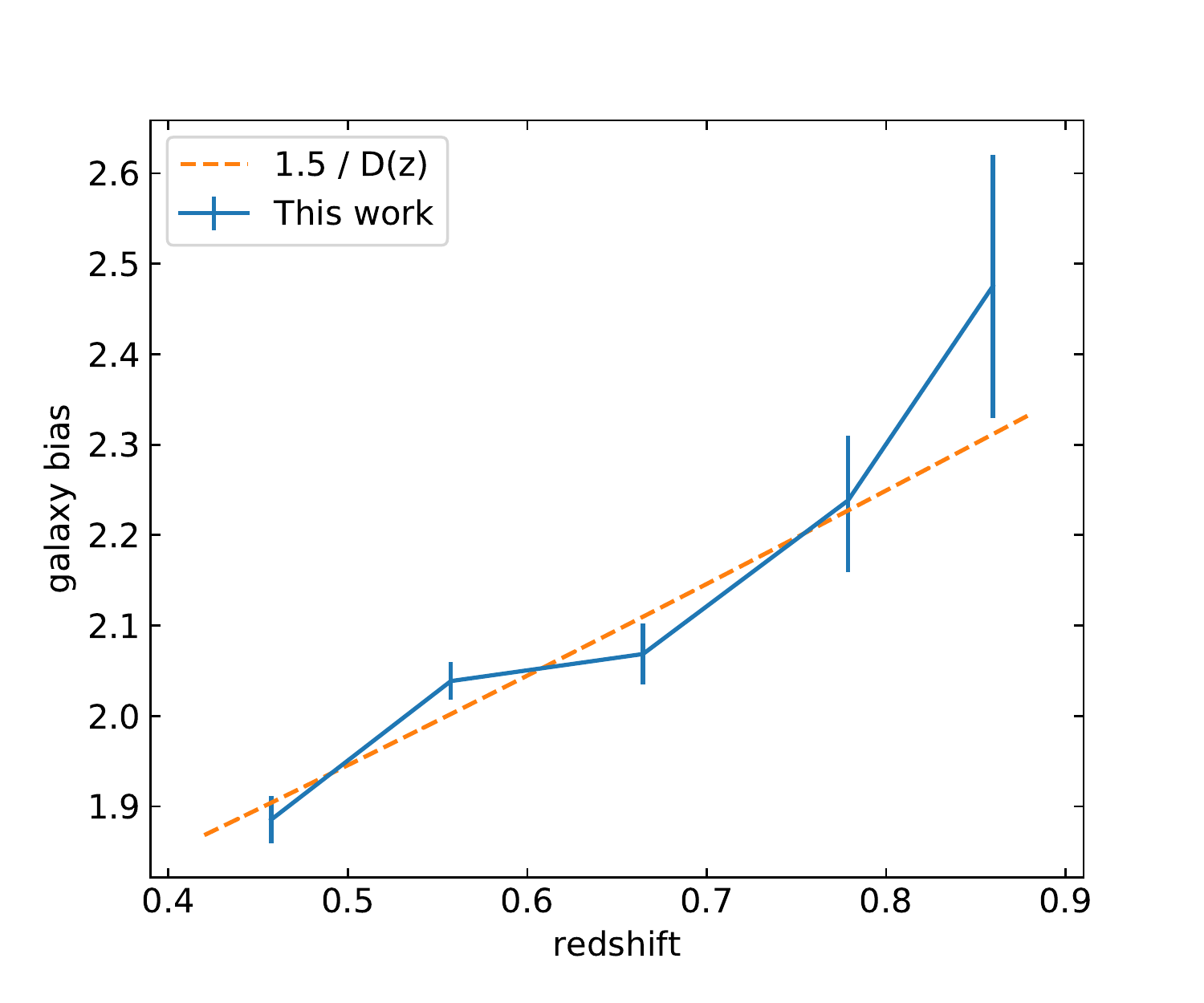}
    \centering
    \caption{The evolution of galaxy bias for our LRG samples. The error bars show the 16\% and 84\% percentiles. The trend is consistent with the bias evolution that would be obtained if one assumed constant clustering amplitude, as shown in the dashed line.}
    \label{fig:bias_evolution}
\end{figure}

The evolution of galaxy bias with redshift, shown in Fig. \ref{fig:bias_evolution}, is consistent with the galaxy bias evolution of a sample with constant clustering amplitude, and it can be written as $b(z) = 1.5/D(z)$, where $D(z)$ is the linear growth factor.
The factor 1.5 is slightly smaller than the factor 1.7 assumed in the DESI Final Design Report \citep{desi_collaboration_desi_2016}.

The BOSS CMASS sample \citep{dawson_baryon_2013}, which has a median redshift of $z \simeq 0.55$, was selected with similarly-motivated luminosity threshold cuts to yield roughly half the comoving number density of our LRG sample. At $z=0.55$, our galaxy bias estimates is consistent with $b \simeq 2.0$ from \citet{white_clustering_2011} for the CMASS sample.
The satellite fraction of our DESI-like LRGs is roughly 15\%, compared to 10\% for CMASS. The fact that there are much fewer satellite galaxies than central galaxies in LRG-like samples is a result of the selection cuts: only the most luminous galaxies are selected, and these luminous galaxies are much more likely to be at the centers of dark matter halos.

The literature on HOD analysis for comparable LRGs at higher redshifts is relatively scarce. The eBOSS survey targeted LRGs in the redshift range of $0.6 < z < 0.9$ \citep{prakash_sdss-iv_2016}, and \citet{zhai_clustering_2017} performed an HOD analysis on the combined BOSS+eBOSS sample in this redshift range. However, the eBOSS LRG sample is significantly different from the our DESI-like LRGs in certain aspects: 1) the comoving number density of the DESI-like LRGs is more than 5 times that of the eBOSS LRGs, and 2) the eBOSS LRG selection does not contain luminosity-threshold cuts, resulting in a wider range of luminosity.
Therefore one should not expect the two samples to have the same HOD or derived parameters. The galaxy bias of DESI-like LRGs at $z{\sim}\,0.7$ (median redshift of eBOSS) is ${\sim}\,2.15$, compared to $2.3$ for eBOSS LRGs. The DESI-like LRGs have a satellite fraction similar to the 13\% for eBOSS LRGs. The DESI-like LRGs have much smaller scatter in the halo mass threshold ($\sigma_{\log M}$) compared to eBOSS (which has $\sigma_{\log M}=0.82$). Lower values of $\sigma_{\log M}$ for the our LRG sample is expected since the sample selection includes a luminosity-threshold cut (whereas the eBOSS LRG selection does not), but as we discussed earlier, our estimate might suffer from inflexibility in the HOD model or photo-$z$ systematics.

The $\alpha$ parameter shows little variation with redshift and is slightly larger than unity; this value is roughly consistent with SDSS \citep{zehavi_galaxy_2011,zentner_constraints_2019} and BOSS \citep{white_clustering_2011} results. It is significantly larger than the value from \citet{zhai_clustering_2017} which estimated $\alpha \sim 0.4$, although this difference could be attributed to the aforementioned differences in the sample selection.

\begin{table*}
    \footnotesize
    \setlength\tabcolsep{2pt}
    \begin{center}
    \caption{Results from the HOD fitting with MCMC for the five redshift bins. For each redshift bin, the first row lists the mean values and the 16th and 84th percentiles; the second row (in italics font) lists the median values; and the third row (in bold font) lists the best-fit values in the six-dimensional HOD parameter space (these need not match the peak of each marginalized posterior distribution).
The parameters $\mathrm{log}(M_{\min})$, $\sigma_{\log M}$, $\alpha$, $\mathrm{log}(M_0)$ and $\mathrm{log}(M_1)$ are free parameters. The bias $b$ and satellite fraction $f_\mathrm{sat}$ are derived parameters. We also list the $\chi^2$ for the best-fit parameters and the corresponding one-sided $p$-value; the fit quality is acceptable for all five redshift bins (values of $p < 0.05$ would indicate statistically significant differences between the model and the data).}
    \label{tab:posterior}
    \begin{tabular}{cccccccccc}
    
        \hline
        Redshift & $\mathrm{log}(M_{\min})$ & $\sigma_{\log M}$ & $\alpha$ & $\mathrm{log}(M_0)$ & $\mathrm{log}(M_1)$ & $b$ & $f_{\mathrm{sat}}$ & $\chi^2$ ($p$-value) \\ \hline
         & $12.90^{+0.05}_{-0.05}$ & $0.17^{+0.13}_{-0.13}$ & $1.28^{+0.09}_{-0.10}$ & $12.13^{+0.64}_{-0.71}$ & $13.99^{+0.05}_{-0.05}$ & $1.89^{+0.03}_{-0.03}$ & $0.15^{+0.02}_{-0.02}$ & \\
        $0.41<z_{\mathrm{phot}}<0.50$ & \textit{12.89} & \textit{0.16} & \textit{1.31} & \textit{12.19} & \textit{14.00} & \textit{1.88} & \textit{0.15} & \\
         & \textbf{12.85} & \textbf{0.03} & \textbf{1.29} & \textbf{12.41} & \textbf{13.97} & \textbf{1.89} & \textbf{0.15} & \textbf{4.57 (0.71)} \\
        \hline
         & $12.92^{+0.03}_{-0.04}$ & $0.12^{+0.09}_{-0.09}$ & $1.11^{+0.22}_{-0.21}$ & $12.73^{+0.40}_{-0.46}$ & $13.98^{+0.07}_{-0.07}$ & $2.04^{+0.02}_{-0.02}$ & $0.13^{+0.01}_{-0.01}$ & \\
        $0.50<z_{\mathrm{phot}}<0.61$ & \textit{12.92} & \textit{0.11} & \textit{1.14} & \textit{12.88} & \textit{13.98} & \textit{2.04} & \textit{0.13} & \\
         & \textbf{12.89} & \textbf{0.00} & \textbf{1.10} & \textbf{12.92} & \textbf{13.95} & \textbf{2.04} & \textbf{0.13} & \textbf{9.87 (0.20)} \\
        \hline
         & $12.86^{+0.05}_{-0.05}$ & $0.20^{+0.13}_{-0.13}$ & $1.31^{+0.06}_{-0.06}$ & $11.94^{+0.58}_{-0.62}$ & $13.92^{+0.04}_{-0.04}$ & $2.07^{+0.03}_{-0.03}$ & $0.16^{+0.01}_{-0.01}$ & \\
        $0.61<z_{\mathrm{phot}}<0.72$ & \textit{12.85} & \textit{0.19} & \textit{1.33} & \textit{11.97} & \textit{13.92} & \textit{2.07} & \textit{0.16} & \\
         & \textbf{12.86} & \textbf{0.28} & \textbf{1.32} & \textbf{11.96} & \textbf{13.91} & \textbf{2.04} & \textbf{0.16} & \textbf{6.86 (0.44)} \\
        \hline
         & $12.97^{+0.08}_{-0.08}$ & $0.29^{+0.19}_{-0.20}$ & $1.28^{+0.12}_{-0.12}$ & $12.05^{+0.69}_{-0.71}$ & $14.00^{+0.05}_{-0.05}$ & $2.24^{+0.07}_{-0.08}$ & $0.14^{+0.02}_{-0.02}$ & \\
        $0.75<z_{\mathrm{phot}}<0.83$ & \textit{12.95} & \textit{0.28} & \textit{1.33} & \textit{12.06} & \textit{14.00} & \textit{2.24} & \textit{0.14} & \\
         & \textbf{12.90} & \textbf{0.02} & \textbf{1.38} & \textbf{12.18} & \textbf{13.98} & \textbf{2.31} & \textbf{0.14} & \textbf{3.29 (0.86)} \\
        \hline
         & $13.26^{+0.13}_{-0.13}$ & $0.45^{+0.20}_{-0.22}$ & $1.20^{+0.31}_{-0.42}$ & $12.33^{+0.91}_{-0.90}$ & $14.31^{+0.07}_{-0.12}$ & $2.48^{+0.15}_{-0.15}$ & $0.10^{+0.02}_{-0.02}$ & \\
        $0.84<z_{\mathrm{phot}}<0.93$ & \textit{13.24} & \textit{0.48} & \textit{1.34} & \textit{12.34} & \textit{14.25} & \textit{2.47} & \textit{0.10} & \\
         & \textbf{13.40} & \textbf{0.70} & \textbf{1.38} & \textbf{11.70} & \textbf{14.24} & \textbf{2.28} & \textbf{0.10} & \textbf{6.65 (0.47)} \\
        \hline

    \end{tabular}
    \end{center}
\end{table*}

Since the HOD model is probabilistic, even if the HOD parameters are fixed, each HOD realization yields a different set of mock galaxies and thus slightly different clustering statistics. This effectively adds a noise to the likelihood function in the MCMC. So long as this noise has mean of zero, it can be shown that as the number of steps becomes large the distribution of points in the chain should still converge to the correct posterior.
However, this ``realization noise'' does cause the likelihoods associated with each step of the MCMC chain to have values which are biased high (or, equivalently, $\chi^2$ to be biased low), since each ``walker'' is less likely to move away from a point whose likelihood was evaluated to be higher than average, and more likely to move away from one which fluctuated low. We show the impact of this bias in Appendix \ref{sec:noisy_mcmc}.
When the realization noise is significant (as here), one cannot directly use the likelihood values from the chain to find the best-fit point or to assess its $\chi^2$; instead, it is necessary to average over repeated realizations of the same model parameters. This could be done at every step of the chain to reduce realization noise, but that is computationally expensive; instead, we adopt an alternative approach.

Specifically, we can exploit the fact that even though the likelihood value associated with each point in the chain is noisy and biased, the set of positions in parameter space that make up the chain do converge to match sampling from the posterior distribution. As a result, the density of points in the chain is highest where the posterior probability is greatest, even when the likelihood values assigned to those points may be inaccurate. We therefore select a small set of steps from the chains which lie in the highest density region of the parameter space; this set is highly likely to contain the points in the chain closest to the best-fit parameters. We then compute the $\chi^2$ values for each of these sets of parameters, averaging over a large number of realizations; from this we can find the point in the chains which truly has the highest likelihood. We describe this procedure in more detail in Appendix \ref{sec:noisy_mcmc}.

Table \ref{tab:posterior} lists the best-fit parameters and the corresponding averaged $\chi^2$ values averaged over 1000 realizations. The table also lists the one-sided $p$-values corresponding to each $\chi^2$ value; i.e., the probability of observing a $\chi^2$ larger than the observed value purely by chance (if we find $p < 0.05$, the hypothesis that the best-fit HOD model matches the data should be rejected). We compute this $p$-value using the number of degrees of freedom $N_{\mathrm{dof}} = N_\mathrm{data} - N_\mathrm{param} = 13 - 6 = 7$, where $N_\mathrm{data}$ is the number of $r_\mathrm{p}$ bins plus one additional constraint from the comoving number density, and $N_\mathrm{param}$ is the number of free parameters. In every case, the HOD model returns a satisfactory fit.

\section{Discussion and conclusion}
\label{sec:conclusion}

We have made a number of methodological improvements for galaxy clustering analysis with the HOD model using photo-$z$'s. We have developed a method that divides the irregular survey footprint into uniform subregions that allowed us to apply the jackknife resampling technique on this dataset. Our methods of correlation measurements using the projected correlation function and the ``cross-correlation'' L-S estimator recover many galaxy pairs straddling the boundaries of the redshift bins, and prevent the counting of pairs which are too far apart in photo-$z$ to have significant clustering but are still placed in the same redshift bin. Both effects boost the S/N in clustering measurements compared to a purely angular clustering analysis. The methods also allow for straightforward and consistent modeling by assigning photo-$z$ errors from the estimated error distribution to the mock galaxies.


With these aforementioned improvements, we have demonstrated that it is possible to obtain good constraints on HOD parameters using only photometric data. Specifically, we have measured the clustering and performed an HOD analysis for DESI LRG target galaxies. We have found that the LRGs are found in massive halos (and especially so for high-z LRGs); this is expected since these are massive and red galaxies, which are only found in the densest environments (e.g., see \citealt{blanton_physical_2009}). We have also found that the host halo properties are very similar for all except the highest-redshift bin. The galaxy bias steadily increases with redshift, increasing from $b \simeq 1.9$ at $z\simeq0.45$ to $b \simeq 2.3$ at $z\simeq0.9$. This trend can be approximated by $b = 1.5/D(z)$, implying constant clustering amplitude over time. The fits prefer a relatively small scatter in the halo mass threshold, suggesting that the LRG selection is efficient in selecting galaxies in massive dark matter halos. At high redshift, the host halos are significantly more massive; this is due to the selection effect of the apparent magnitude limit on the galaxy sample.

The results of this paper can be used to create improved mock galaxy catalogs for DESI. The upcoming spectroscopic data from DESI will eliminate the uncertainties from photo-$z$'s and provide tests of our results, although there will instead be systematics from fiber collisions to be dealt with. The spectroscopic redshifts will also enable the accurate measurements of the rest-frame colors and luminosity, and it would be interesting to study the color and luminosity dependence of the galaxy clustering; results from such studies can provide important insights into the formation and evolution of these massive galaxies.

The overall methodology of HOD modeling with photo-$z$'s presented here can be easily implemented with existing analysis codes such as \textsc{halotools}, and it can be adopted for future imaging surveys such as LSST for studying the galaxy-halo connection. There are several aspects where our methods can be further improved upon by adding more sophistication, which we discuss below.


First, the approximation of Gaussian photo-$z$ errors is not always appropriate. In our case, the LRGs have prominent spectral features such as the $4000\text{\AA}$ break and the $1.6$ micron bump that result in unambiguous photo-$z$'s. Therefore, we are able to treat each PDF as a simple Gaussian distribution, and to assume that the photo-$z$ errors are dominated by photometric uncertainties; these simplifications are supported by spectroscopic validation. However, in many other cases, the Gaussian approximation is not sufficient, and one needs to take the full photo-$z$ PDFs as input in the fitting process; this can be important for galaxies that have skewed or multimodal PDFs or for datasets that have weak constraining power on redshifts.

Second, although in our case the uncertainty in the calibration of photo-$z$ errors (quantified by $S_z$) is subdominant at lower redshifts, at higher redshift it causes significant uncertainties in HOD parameters and galaxy bias. Therefore, in the presence of relatively large photo-$z$ errors, better priors on the calibration of photo-$z$ uncertainties would significantly reduce the uncertainties in the model parameters. This can be achieve using a small spectroscopic subsample that is representative of the full photometric sample.

Third, in this work we have assumed that the our galaxy sample has the same intrinsic photo-$z$ error properties, and the distribution of actual photo-$z$ errors are solely due to the variation in the S/N of the photometry; therefore we can randomly draw from the estimated photo-$z$ errors and assign the resulting values to each mock galaxy. This assumption might not hold for a sample of galaxies that are more diverse than the LRGs; for example, in a pure luminosity threshold sample, the blue and red galaxies will have very different intrinsic photo-$z$ errors/PDFs. Nevertheless, the method can account for such differences by treating differently halos with different properties when assigning photo-$z$ errors to the corresponding galaxies.

Finally, since the correlation function is measured with relative distances, it is insensitive to an overall offset in photo-$z$'s so long as the offset is the same for all galaxies in the sample. Our model does not account for higher order offsets, and light-cone mocks would be required to simulate such effects.

With these improved methods and enlarged samples from future surveys, fully photometric HOD modeling will be a powerful tool for studying the galaxy-halo connection with future imaging surveys such as LSST.

\section*{Acknowledgements}


The authors would like to thank Hee-Jong Seo, Jeremy Tinker and Gustavo Niz for their feedback on the draft and useful discussions. RZ and JAN were supported by the U.S. Department of Energy Office of Science, Office of High Energy Physics via grant DE-SC0007914. RZ was also supported by the Director, Office of Science, Office of High Energy Physics of the U.S. Department of Energy under Contract No. DE-AC02-05CH1123. Support for YYM was provided by the Pittsburgh Particle Physics, Astrophysics and Cosmology Center through the Samuel P.\ Langley PITT PACC Postdoctoral Fellowship, and by NASA through the NASA Hubble Fellowship grant no.\ HST-HF2-51441.001 awarded by the Space Telescope Science Institute, which is operated by the Association of Universities for Research in Astronomy, Incorporated, under NASA contract NAS5-26555. JM gratefully acknowledges support from NSF grant AST- 1616414 and DOE grant DE-SC0020086. ADM was supported by the U.S.Department of Energy, Office of Science, Office of High Energy Physics, under Award Number DE-SC0019022. ARZ was funded by the US National Science Foundation (NSF) through grants AST 1516266 and AST 1517563. DYT thanks Prof. Steve Ahlen for his mentorship and support and ackowledges the generous support by U.S. Department of Energy Office of Science, grant No. DE-SC0015628.

This research is supported by the Director, Office of Science, Office of High Energy Physics of the U.S. Department of Energy under Contract No. DE–AC02–05CH1123, and by the National Energy Research Scientific Computing Center, a DOE Office of Science User Facility under the same contract; additional support for DESI is provided by the U.S. National Science Foundation, Division of Astronomical Sciences under Contract No. AST-0950945 to the National Optical Astronomy Observatory; the Science and Technologies Facilities Council of the United Kingdom; the Gordon and Betty Moore Foundation; the Heising-Simons Foundation; the French Alternative Energies and Atomic Energy Commission (CEA); the National Council of Science and Technology of Mexico, and by the DESI Member Institutions. The authors are honored to be permitted to conduct astronomical research on Iolkam Du’ag (Kitt Peak), a mountain with particular significance to the Tohono O’odham Nation.

The Legacy Surveys consist of three individual and complementary projects: the Dark Energy Camera Legacy Survey (DECaLS; NOAO Proposal ID \# 2014B-0404; PIs: David Schlegel and Arjun Dey), the Beijing-Arizona Sky Survey (BASS; NOAO Proposal ID \# 2015A-0801; PIs: Zhou Xu and Xiaohui Fan), and the Mayall z-band Legacy Survey (MzLS; NOAO Proposal ID \# 2016A-0453; PI: Arjun Dey). DECaLS, BASS and MzLS together include data obtained, respectively, at the Blanco telescope, Cerro Tololo Inter-American Observatory, National Optical Astronomy Observatory (NOAO); the Bok telescope, Steward Observatory, University of Arizona; and the Mayall telescope, Kitt Peak National Observatory, NOAO. The Legacy Surveys project is honored to be permitted to conduct astronomical research on Iolkam Du'ag (Kitt Peak), a mountain with particular significance to the Tohono O'odham Nation.

NOAO is operated by the Association of Universities for Research in Astronomy (AURA) under a cooperative agreement with the National Science Foundation.

This project used data obtained with the Dark Energy Camera (DECam), which was constructed by the Dark Energy Survey (DES) collaboration. Funding for the DES Projects has been provided by the U.S. Department of Energy, the U.S. National Science Foundation, the Ministry of Science and Education of Spain, the Science and Technology Facilities Council of the United Kingdom, the Higher Education Funding Council for England, the National Center for Supercomputing Applications at the University of Illinois at Urbana-Champaign, the Kavli Institute of Cosmological Physics at the University of Chicago, Center for Cosmology and Astro-Particle Physics at the Ohio State University, the Mitchell Institute for Fundamental Physics and Astronomy at Texas A\&M University, Financiadora de Estudos e Projetos, Fundacao Carlos Chagas Filho de Amparo, Financiadora de Estudos e Projetos, Fundacao Carlos Chagas Filho de Amparo a Pesquisa do Estado do Rio de Janeiro, Conselho Nacional de Desenvolvimento Cientifico e Tecnologico and the Ministerio da Ciencia, Tecnologia e Inovacao, the Deutsche Forschungsgemeinschaft and the Collaborating Institutions in the Dark Energy Survey. The Collaborating Institutions are Argonne National Laboratory, the University of California at Santa Cruz, the University of Cambridge, Centro de Investigaciones Energeticas, Medioambientales y Tecnologicas-Madrid, the University of Chicago, University College London, the DES-Brazil Consortium, the University of Edinburgh, the Eidgenossische Technische Hochschule (ETH) Zurich, Fermi National Accelerator Laboratory, the University of Illinois at Urbana-Champaign, the Institut de Ciencies de l'Espai (IEEC/CSIC), the Institut de Fisica d'Altes Energies, Lawrence Berkeley National Laboratory, the Ludwig-Maximilians Universitat Munchen and the associated Excellence Cluster Universe, the University of Michigan, the National Optical Astronomy Observatory, the University of Nottingham, the Ohio State University, the University of Pennsylvania, the University of Portsmouth, SLAC National Accelerator Laboratory, Stanford University, the University of Sussex, and Texas A\&M University.

BASS is a key project of the Telescope Access Program (TAP), which has been funded by the National Astronomical Observatories of China, the Chinese Academy of Sciences (the Strategic Priority Research Program "The Emergence of Cosmological Structures" Grant \# XDB09000000), and the Special Fund for Astronomy from the Ministry of Finance. The BASS is also supported by the External Cooperation Program of Chinese Academy of Sciences (Grant \# 114A11KYSB20160057), and Chinese National Natural Science Foundation (Grant \# 11433005). The Legacy Survey team makes use of data products from the Near-Earth Object Wide-field Infrared Survey Explorer (NEOWISE), which is a project of the Jet Propulsion Laboratory/California Institute of Technology. NEOWISE is funded by the National Aeronautics and Space Administration.

The Legacy Surveys imaging of the DESI footprint is supported by the Director, Office of Science, Office of High Energy Physics of the U.S. Department of Energy under Contract No. DE-AC02-05CH1123, by the National Energy Research Scientific Computing Center, a DOE Office of Science User Facility under the same contract; and by the U.S. National Science Foundation, Division of Astronomical Sciences under Contract No. AST-0950945 to NOAO.

Funding for the DEEP2 Galaxy Redshift Survey has been provided by NSF grants AST-95-09298, AST-0071048, AST-0507428, and AST-0507483 as well as NASA LTSA grant NNG04GC89G. Funding for the DEEP3 Galaxy Redshift Survey has been provided by NSF grants AST-0808133, AST-0807630, and AST-0806732. Funding for the SDSS and SDSS-II has been provided by the Alfred P. Sloan Foundation, the Participating Institutions, the National Science Foundation, the U.S. Department of Energy, the National Aeronautics and Space Administration, the Japanese Monbukagakusho, the Max Planck Society, and the Higher Education Funding Council for England. Funding for SDSS-III has been provided by the Alfred P. Sloan Foundation, the Participating Institutions, the National Science Foundation, and the U.S. Department of Energy Office of Science. Funding for the Sloan Digital Sky Survey IV has been provided by the Alfred P. Sloan Foundation, the U.S. Department of Energy Office of Science, and the Participating Institutions. SDSS-IV acknowledges support and resources from the Center for High-Performance Computing at the University of Utah. GAMA is a joint European-Australasian project based around a spectroscopic campaign using the Anglo-Australian Telescope. The GAMA input catalogue is based on data taken from the Sloan Digital Sky Survey and the UKIRT Infrared Deep Sky Survey. Complementary imaging of the GAMA regions is being obtained by a number of independent survey programmes including GALEX MIS, VST KiDS, VISTA VIKING, WISE, Herschel-ATLAS, GMRT and ASKAP providing UV to radio coverage. GAMA is funded by the STFC (UK), the ARC (Australia), the AAO, and the participating institutions. The GAMA website is http://www.gama-survey.org/. This paper uses data from the VIMOS Public Extragalactic Redshift Survey (VIPERS). VIPERS has been performed using the ESO Very Large Telescope, under the ``Large Programme'' 182.A-0886. The participating institutions and funding agencies are listed at http://vipers.inaf.it. This research uses data from the VIMOS VLT Deep Survey, obtained from the VVDS database operated by Cesam, Laboratoire d'Astrophysique de Marseille, France.





\appendix

\section{WISE bright star masks}
\label{sec:app_masks}

\begin{figure}
    \includegraphics[width=0.95\columnwidth]{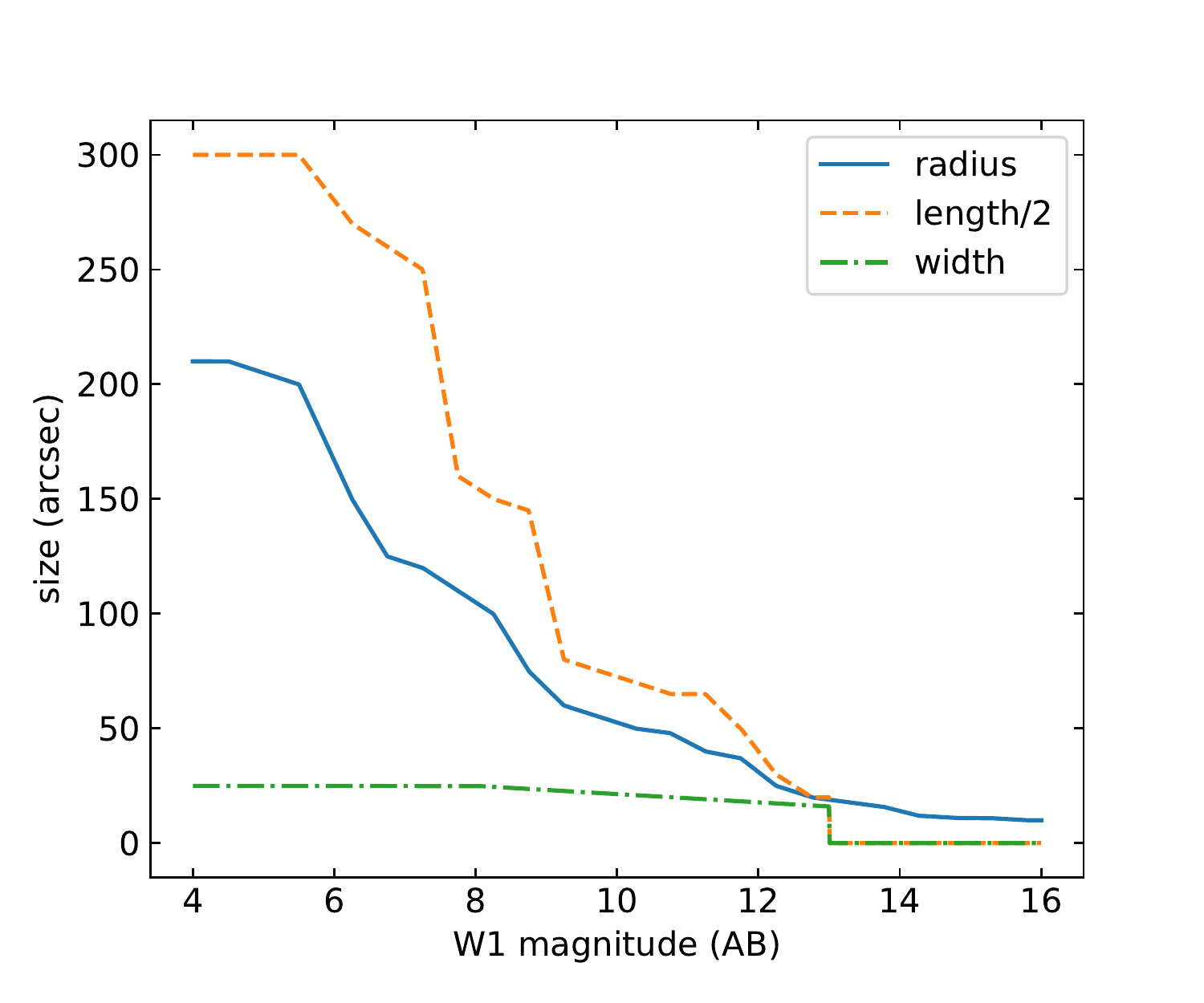}
    \centering
    \caption{The circular mask radius and the length and width of the rectangular mask as a function of W1 magnitude of the bright star. No masking of diffraction spikes is performed for stars fainter than $W1_{\mathrm{AB}}=13.0$.}
    \label{fig:mask_size}
\end{figure}

\begin{figure*}
    \centering
    \begin{subfigure}[b]{0.4\textwidth}
        \includegraphics[width=\textwidth]{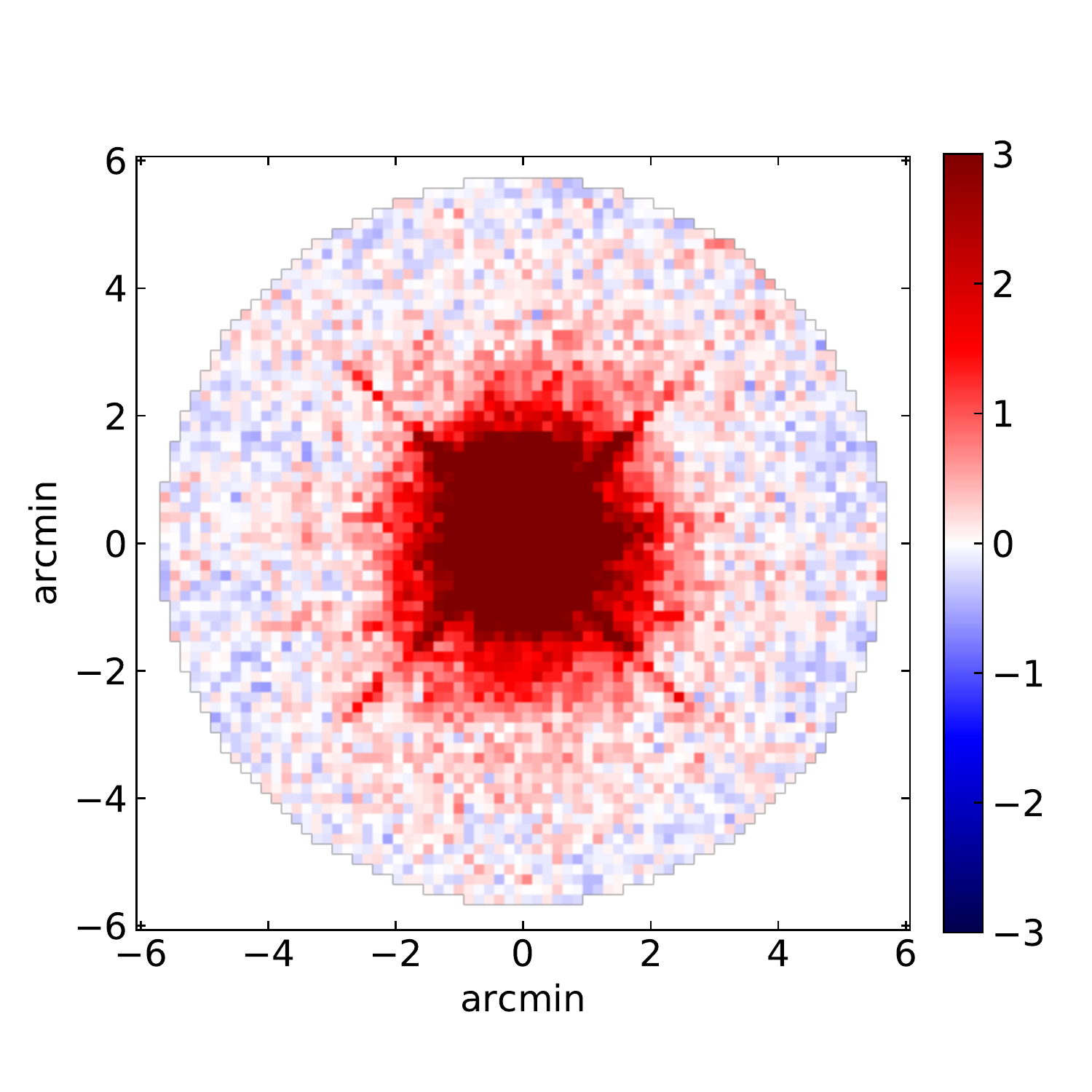}
    \end{subfigure}
    \begin{subfigure}[b]{0.4\textwidth}
        \includegraphics[width=\textwidth]{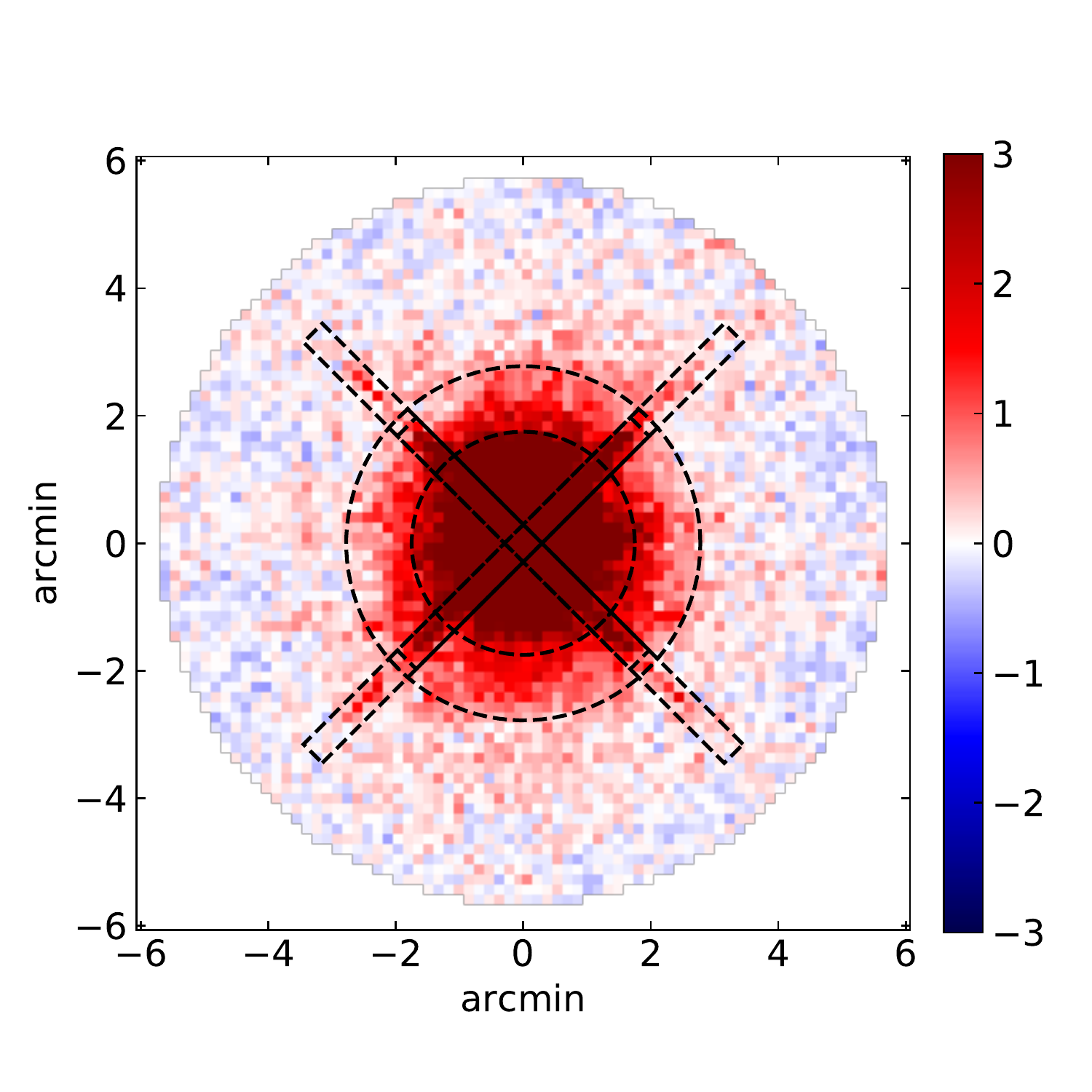}
    \end{subfigure}
    \caption{Left panel: Cross-correlation between LRGs and AllWISE bright stars with $6<W1_{\mathrm{AB}}<8$ in ecliptic coordinates. The colors represent the fractional over/under-density of LRGs compared to the overall average density (as a result the values cannot drop below $-1$). Right panel: the boundaries of the WISE masks at $W1_{\mathrm{AB}}=6$ and $W1_{\mathrm{AB}}=8$ are overplotted.}
    \label{fig:wise_crosscorr}
\end{figure*}

This appendix provides some technical details of the WISE bright star masks that we described in section \ref{sec:mask}.
The WISE bright star masks are geometric masks around stars in the AllWISE catalog with $W1_{AB}<16.0$. The masks consist of two components: circular masks for the ``core'' of a bright star, and rectangular masks for the diffraction spikes. The size of the masks vary with the $W1$ magnitude of the bright star, and the size-magnitude relation is shown in Fig. \ref{fig:mask_size}. To obtain the size-magnitude relation, we cross-correlate the positions on the sky between LRGs and bright stars in magnitude bins of $\Delta W1_{AB}=0.5$, and locate (by hand) the distance where the LRG density starts to noticeably deviate from the densities further away from the stars. This way the majority of the contaminated objects in the LRG sample are masked. Fig. \ref{fig:wise_crosscorr} shows the cross-correlation between the LRG sample and AllWISE stars with $6<W1_{\mathrm{AB}}<8$ with the masks overlaid.

\section{Photometric Redshifts for Legacy Surveys Data Release 8}
\label{sec:dr8_pz}

Accompanying this paper we have released an updated version of the photometric redshifts used to conduct the large-scale-structure analyses we have presented. This constitutes the Photometric Redshifts for the Legacy Surveys (PRLS) catalog; it is built using the most recent DR8 dataset of the Legacy Surveys\footnote{\url{https://www.legacysurvey.org/dr8/}}. Data Release 8 includes data from both the DECaLS survey in the southern portion of the planned DESI footprint ($\mathrm{DEC}\leqslant34\degr$) and the BASS and MzLS surveys in the northern sky ($\mathrm{DEC}\geqslant32\degr$). The two regions have slightly different effective response curves due to variations in filter transmission, detectors (for the $g$ and $r$ bands), etc., and thus photo-$z$ algorithms for each must be trained separately. Table \ref{tab:dr8_truth_catalogs} lists the redshift surveys used as training samples in each region. In addition to the redshift catalogs  used for training the DR7 photo-$z$'s (described in \S \ref{sec:truth}), in the BASS/MzLS region we can also utilize redshifts from the DEEP3 Galaxy Redshift Survey \citep{cooper_deep3_2012,zhou_deep_2019}. The method for computing the DR8 photo-$z$'s is the same as described in section \ref{sec:lrg_photo-z}; as a result the photo-$z$ performance in DR8 South is very similar to the DR7 photo-$z$'s used in this paper. 

The columns of the PRLS photo-$z$ catalog are described in Table \ref{tab:pz_catalog_columns}.

\begin{table}
    \centering
    \caption{Number of objects from each redshift survey that are cross-matched to Legacy Surveys DR8 (before downsampling). For datasets that have been downsampled for photo-$z$ training, we also list percentages of the remaining objects after downsampling.}
    \label{tab:dr8_truth_catalogs}
    \begin{tabular}{lll}
        \hline
         Survey                & DR8 North        & DR8 South       \\
        \hline
         2dFLenS               & --               & 28416           \\
         AGES                  & 16197            & 4057            \\
         BOSS                  & 377769  (39.6\%) & 951829 (33.3\%) \\
         COSMOS2015            & --               & 62510           \\
         DEEP2 Field 2, 3, 4   & 14401            & 20602           \\
         DEEP2 Field 1 + DEEP3 & 13355            & --              \\
         GAMA                  & --               & 123293 (52.9\%) \\
         OzDES                 & 244              & 5174            \\
         SDSS                  & 281636 (46.0\%)  & 574767 (43.9\%) \\
         VIPERS                & --               & 47328           \\
         VVDS                  & --               & 6321            \\
         WiggleZ               & 1627             & 144963 (32.6\%) \\
         eBOSS                 & 31330            & 38459           \\
        \hline
    \end{tabular}
\end{table}

To assess the photo-$z$ accuracy for the spectroscopic training sample, we employ 5-fold cross-validation. To do this we randomly divide the dataset into 5 equal chunks; we can then combine 4 chunks for training and evaluate the performance with the remaining chunk. We repeat this procedure until all 5 chunks have been used for testing. In this way the entire truth dataset can be utilized for testing without biasing the assessment of performance.

\begin{figure}
    \includegraphics[width=0.9\columnwidth]{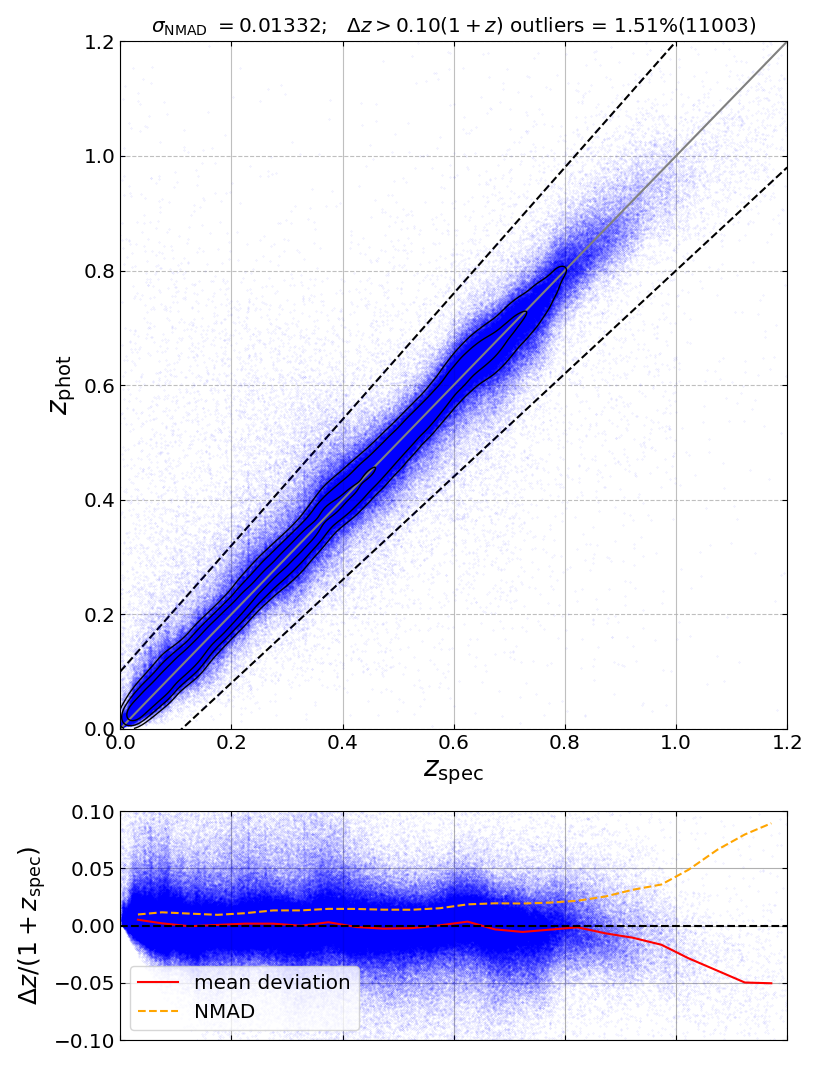}
    \centering
    \caption{Photo-$z$ vs. spec-$z$ plot similar to Fig. \ref{fig:pz_weighted}, but for $z_{\mathrm{mag}}<21.0$ objects with spec-$z$ training objects in DR8 South (the region covered by the DECaLS survey) without any weighting (cross-validation is used to avoid biasing errors low). The photometric redshifts are generally well-behaved in this regime.}
    \label{fig:dr8_south_pz_bright}
\end{figure}

\begin{figure}
    \includegraphics[width=0.9\columnwidth]{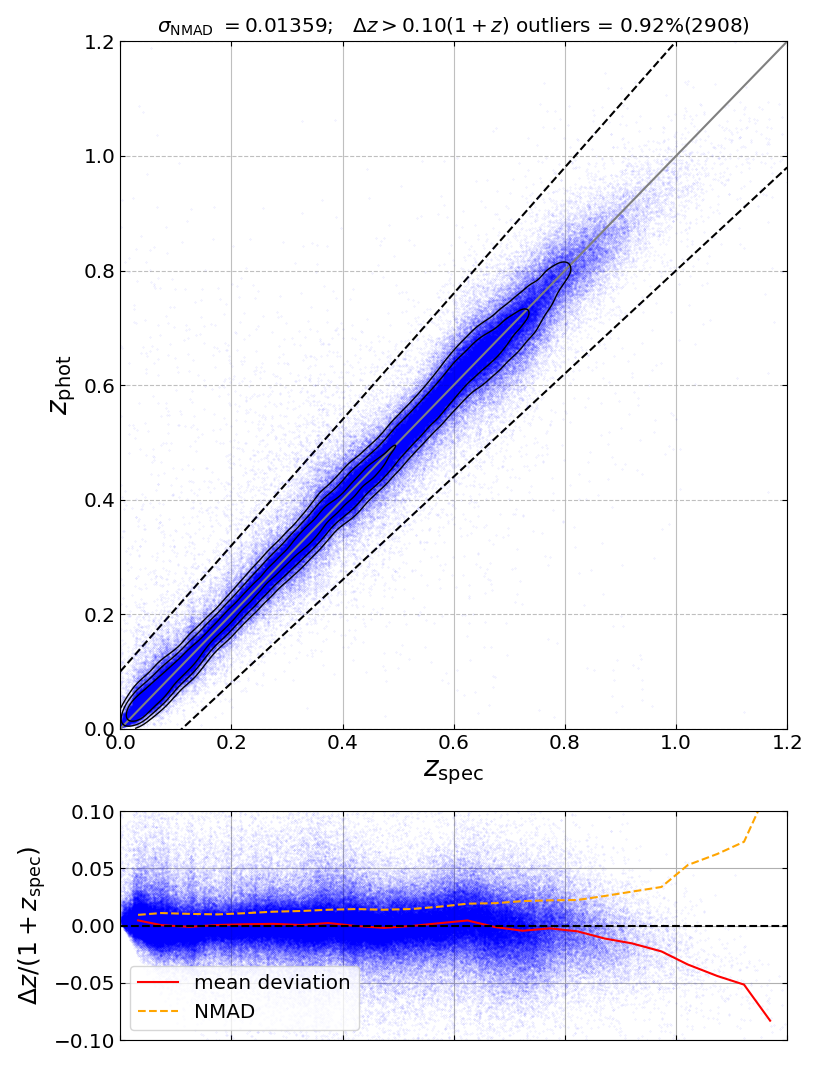}
    \centering
    \caption{As Fig. \ref{fig:dr8_south_pz_bright}, but for DR8 North (the BASS/MzLS survey region).}
    \label{fig:dr8_north_pz_bright}
\end{figure}

Fig. \ref{fig:dr8_south_pz_bright} shows the relationship between photo-$z$ and spec-$z$ for $z_{\mathrm{mag}}<21.0$ objects in the truth catalog in DR8 South (the DECaLS region).  For the (unweighted) objects, the photo-$z$ scatter is ${\sim}\,0.013$ and the outlier rate is $1.5\%$, although it is worth noting that this sample is dominated by bright galaxies from surveys like SDSS and BOSS and therefore the numbers do not represent the photo-$z$ accuracy of a magnitude limited sample with $z_{\mathrm{mag}}<21.0$. Fig. \ref{fig:dr8_north_pz_bright} presents the equivalent plot for DR8 North (the BASS/MzLS region).

\begin{figure}
    \includegraphics[width=0.9\columnwidth]{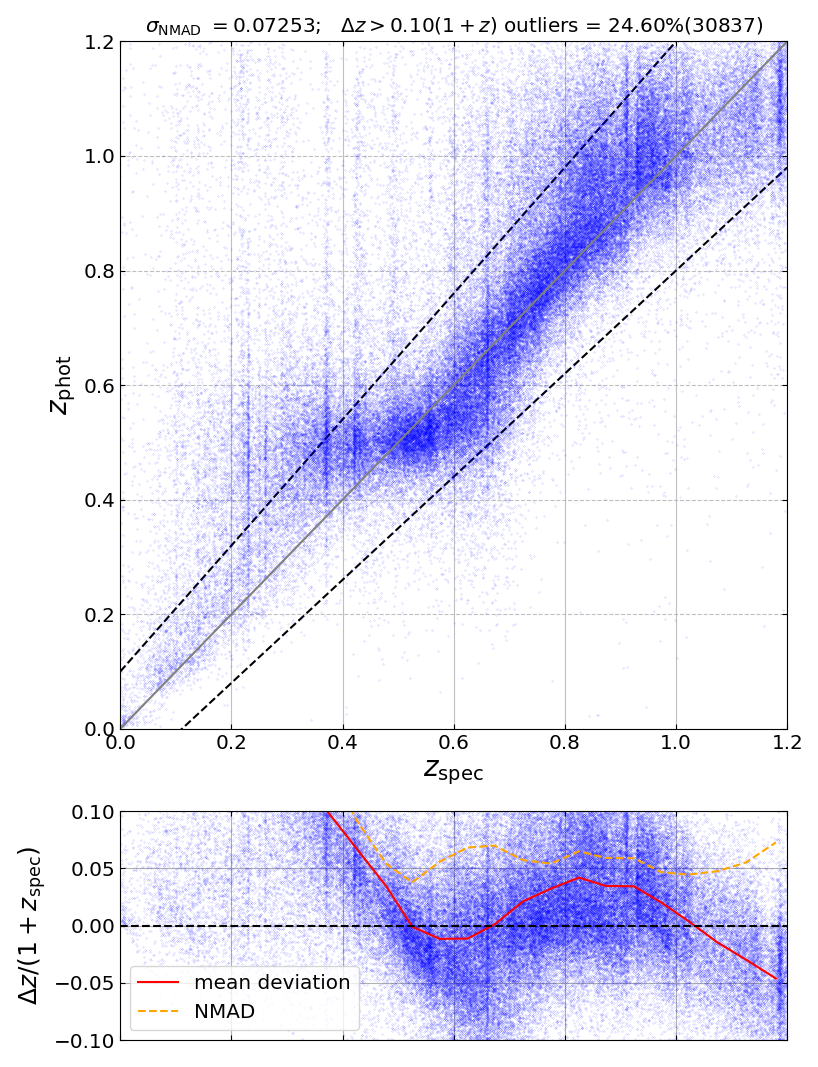}
    \centering
    \caption{Same as Fig. \ref{fig:dr8_south_pz_bright} but for $z_{\mathrm{mag}}>21.0$ objects in the truth catalog. In this regime the photo-$z$'s are poorly constrained, especially for objects with $z \lesssim 0.5$.}
    \label{fig:dr8_south_pz_faint}
\end{figure}

To show how the photo-$z$'s start to systematically break down beyond $z_{\mathrm{mag}} \simeq 21$, we plot photo-$z$ vs spec-$z$ for $z_{\mathrm{mag}}>21.0$ objects (in DR8 South) in Fig. \ref{fig:dr8_south_pz_faint}. The failure of the photo-$z$'s is due to the limitations in our imaging data: $W1$ and $W2$ are too shallow to be useful for the fainter galaxies, so for these galaxies we are effectively limited to only the three optical bands, $grz$, which are not sufficient to constrain the photo-$z$'s. The shallow $W1/W2$ imaging is particularly problematic for galaxies with $z \lesssim 0.5$, as their $1.6$ micron bump is still far from the $W1$ band, causing their photo-$z$'s to fail catastrophically.

\begin{table*}
    \centering
    \caption{Descriptions of the columns in the PRLS photometric redshift catalog.}
    \label{tab:pz_catalog_columns}
    \begin{tabular}{ll}
        \hline
        Name            & Description                                                                \\
        \hline
        z\_phot\_mean   & photo-$z$ derived from the mean of the photo-$z$ PDF                           \\
        z\_phot\_median & photo-$z$ derived from the median of the photo-$z$ PDF                         \\
        z\_phot\_std    & standard deviation of the photo-$z$'s derived from the photo-$z$ PDF             \\
        z\_phot\_l68    & lower bound of the 68\% confidence region, derived from the photo-$z$ PDF                   \\
        z\_phot\_u68    & upper bound of the 68\% confidence region, derived from the photo-$z$ PDF               \\
        z\_phot\_l95    & lower bound of the 95\% confidence region, derived from the photo-$z$ PDF               \\
        z\_phot\_u95    & upper bound of the 68\% confidence region, derived from the photo-$z$ PDF                    \\
        z\_spec         & spectroscopic redshift, if available                                       \\
        survey          & source of the spectroscopic redshift                                       \\
        training        & whether or not the spectroscopic redshift is used in photometric redshift training \\
        \hline
    \end{tabular}
\end{table*}

\section{HOD parameters from skew normal photo-$z$ PDF}
\label{sec:skewnorm}
\begin{figure}
    \includegraphics[width=0.95\columnwidth]{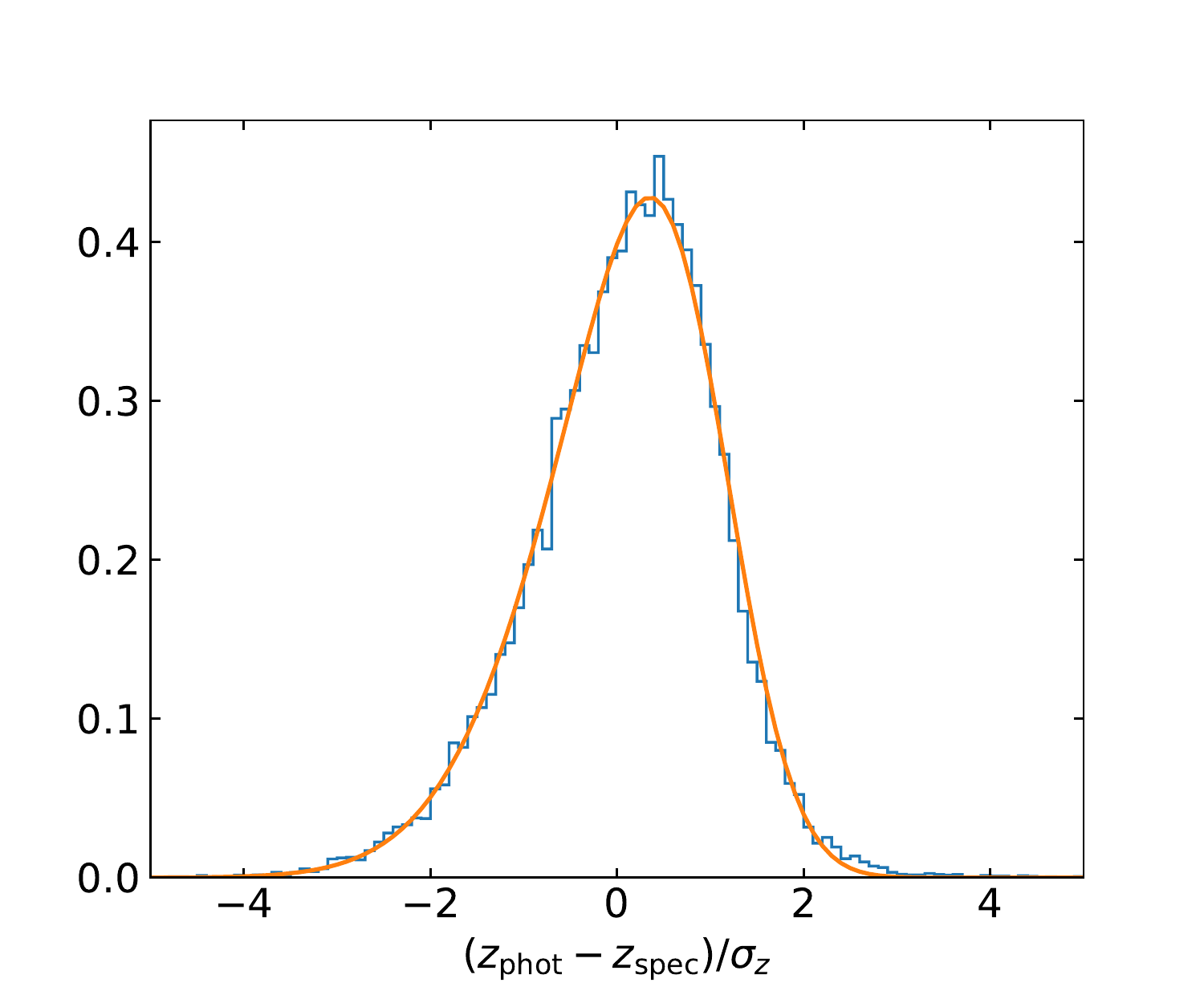}
    \centering
    \caption{Illustration of our fit of a skew normal distribution to the redshift error distribution.   As in Fig. \ref{fig:pz_error_validation}, the histogram shows the distribution of redshift errors for objects with spectroscopic measurements in units of the estimated uncertainty for a given object for the $0.61<z<0.72$ redshift bin. The curve corresponds to the PDF of the best-fit skew normal distribution, with $\alpha=-2.1$.}
    \label{fig:skewnorm}
\end{figure}

The distributions of $(z_{\mathrm{phot}}-z_{\mathrm{spec}})/\sigma_z$  shown in Fig. \ref{fig:pz_error_validation} suggests that the distribution of photo-$z$ errors may be slightly skewed (e.g., for the redshift bin covering $0.61<z<0.72$). To see if such a skewness could cause systematic offsets in our HOD modeling, we have rerun the MCMC for the redshift bin of $0.61<z<0.72$ treating all individual objects' photo-$z$ PDFs as skew normal distributions with the same standard deviation that was used in the Gaussian representation. We obtain the $\alpha$ parameter of the distribution (which determines its skewness) by fitting a skew normal distribution to the $(z_{\mathrm{phot}}-z_{\mathrm{spec}})/\sigma_z$ distribution, as shown in Fig. \ref{fig:skewnorm}. We find the best-fit $\alpha$ value to be -2.1, which we keep fixed in the MCMC, using the same $\alpha$ value to skew the error distributions for each mock object. As in our standard analysis, the photo-$z$ errors are rescaled by a constant factor of $S_z$ which is treated as a nuisance parameter. The resulting HOD parameters are listed in Table \ref{tab:posterior_skewnorm}, and for comparison we also list the results from the our standard analysis (which treats the photo-$z$ PDFs as Gaussian). We find that there are negligible changes in the inferred HOD parameters when switching to a skew normal PDF with the skewness estimated from the available spec-$z$ data and standard deviation set by our error estimates.

\begin{table}
    \footnotesize
    \setlength\tabcolsep{2pt}
    \begin{center}
    \caption{The mean values and the 16th and 84th percentiles of the HOD parameters from the MCMC chain, for the Gaussian photo-$z$ PDF and the skew normal PDF. In each case differences are small compared to the statistical uncertainties.}
    \label{tab:posterior_skewnorm}
    \begin{tabular}{cccccccccc}
        \hline
         & $\mathrm{log}(M_{\min})$ & $\sigma_{\log M}$ & $\alpha$ & $\mathrm{log}(M_0)$ & $\mathrm{log}(M_1)$ \\ \hline
        Gaussian & $12.86^{+0.05}_{-0.05}$ & $0.20^{+0.13}_{-0.13}$ & $1.31^{+0.06}_{-0.06}$ & $11.94^{+0.58}_{-0.62}$ & $13.92^{+0.04}_{-0.04}$ \\
        Skew normal & $12.86^{+0.05}_{-0.05}$ & $0.20^{+0.13}_{-0.14}$ & $1.31^{+0.06}_{-0.07}$ & $11.98^{+0.56}_{-0.61}$ & $13.92^{+0.04}_{-0.04}$ \\
        \hline
    \end{tabular}
    \end{center}
\end{table}

\section{Identifying Maximum-Likelihood Values in Noisy MCMC}
\label{sec:noisy_mcmc}
Here we provide more details on the treatment of noisy MCMC that we discussed in the last part of section \ref{sec:mcmc_results}.

\begin{figure}
    \includegraphics[width=0.95\columnwidth]{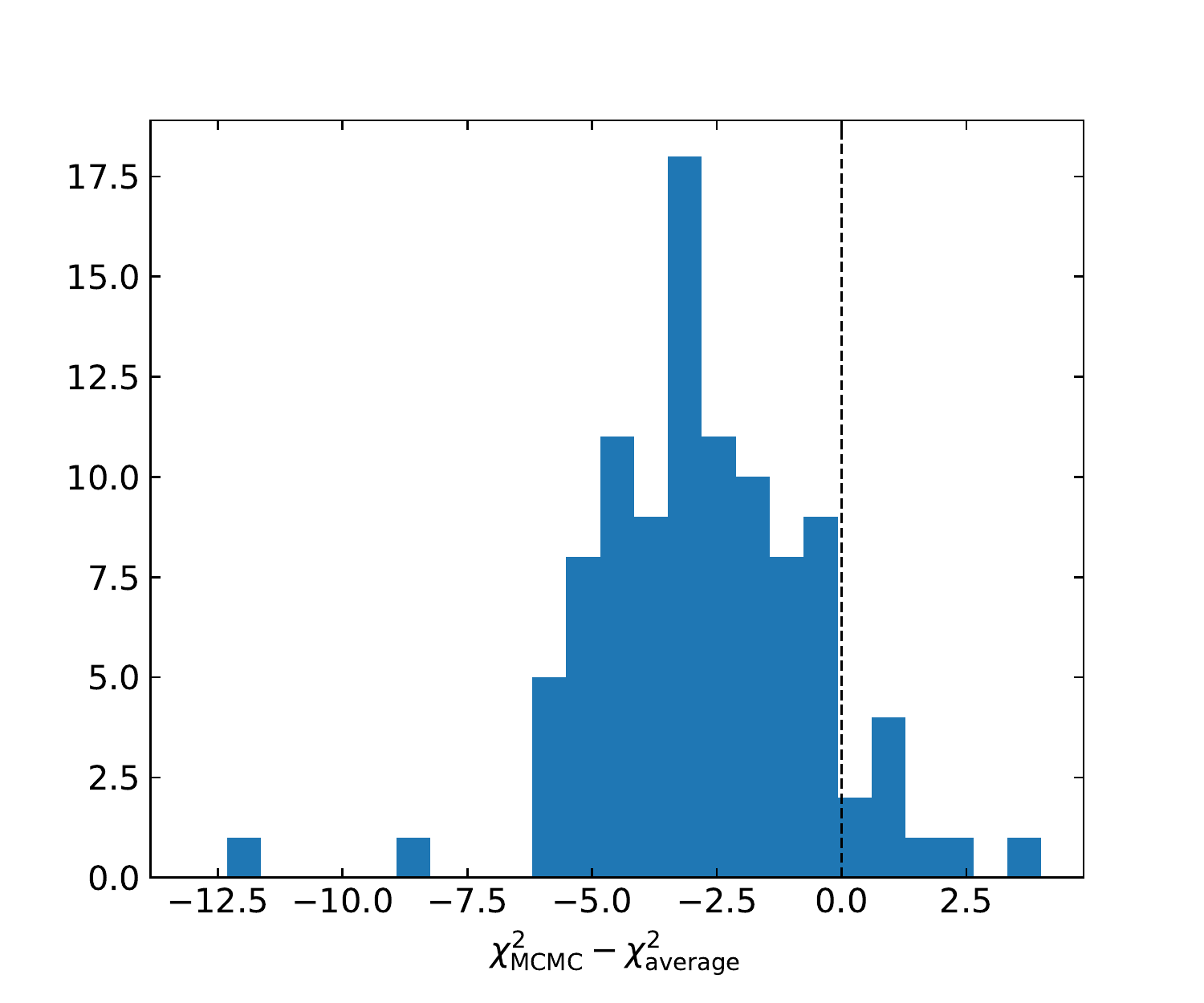}
    \centering
    \caption{Histogram of the difference between the $\chi^2$ value from the MCMC chain and the mean $\chi^2$ value averaging over 100 realizations for the same set of HOD parameters. The vertical line highlights the location of zero difference. The $\chi^2$ difference is clearly non-zero on average, indicating that the $\chi^2$ values from the MCMC chain are biased low.}
    \label{fig:mcmc_chi2_bias}
\end{figure}

First, to demonstrate that the likelihoods (or, equivalently, $\chi^2$ values, since we use flat priors and Gaussian likelihoods) at each step from the MCMC chain are indeed biased, we randomly select 100 points from the chain (where each point corresponds to a set of model parameters), and for each of these points we rerun 100 HOD realizations to obtain the average $\chi^2$ value. Figure \ref{fig:mcmc_chi2_bias} shows the difference between the $\chi^2$ value taken from the MCMC chain and the mean $\chi^2$ value averaging over 100 realizations for the same position in parameter space, drawn from the fits for the redshift bin of $0.61<z<0.72$ (the same redshift bin is used for all other figures in this section). Clearly the $\chi^2$ values from the chain are biased low compared to the averaged values. This bias is caused by noise in the MCMC likelihood function, as discussed in section \ref{sec:mcmc_results}.

\begin{figure*}
   \includegraphics[width=0.97\textwidth]{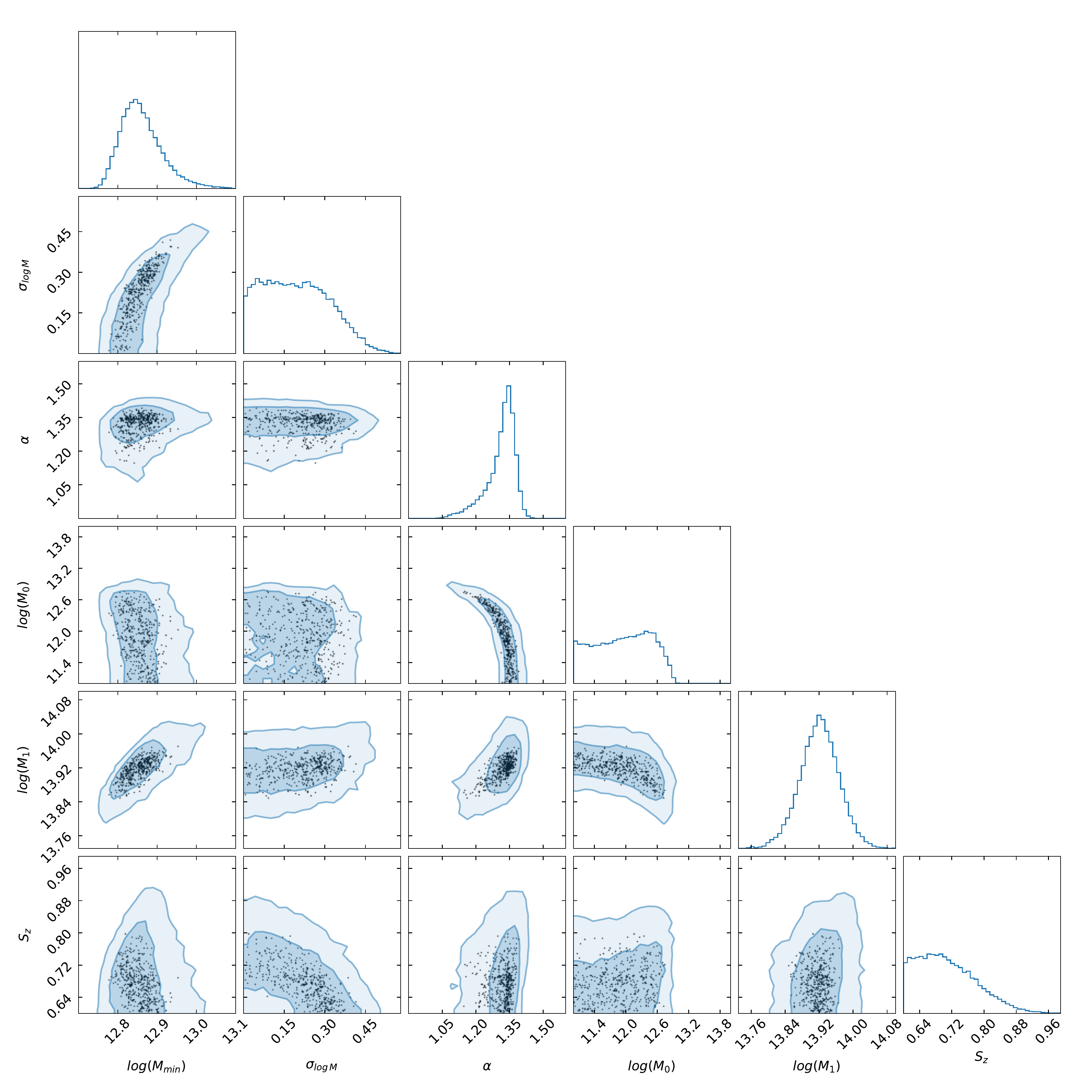}
   \centering
    \caption{Points with the highest chain likelihood values selected from 500 segments in the chain for the redshift bin of $0.61<z<0.72$. The contours of the full MCMC chain are also shown. The highest-likelihood points span a wide range in parameter space due to the stochastic nature of the model likelihood evaluated at a given point in the chain.}
    \label{fig:mcmc_highest_lilelihood}
\end{figure*}

To demonstrate the difficulty of using the likelihood from the MCMC chain for finding the best-fit parameters in the presence of realization noise, here we show that the points selected to have the highest likelihood values have a very wide distribution in the parameter space. In MCMC the chain positions are correlated, and sometimes a walker can get ``stuck'' at the same position for many steps; to reduce such effects, we divide the chain into 500 segments, and select the highest likelihood point in each segment, and plot their positions in parameter space in Fig. \ref{fig:mcmc_highest_lilelihood}. Even though these points have higher likelihoods in the chain than 99.95\% of the sample, they span a large range in parameter space.

As selecting the best-fit parameters using the chain likelihoods does not yield good results, we adopt an alternative approach. The distribution of points in the chain should converge to follow the posterior. As a result, the density of points in the chain should be greatest where the posterior is highest.  We can therefore select points in the regions of highest density and they are likely to be near the likelihood peak. Below we describe the method in detail.

First we downsample the MCMC chain by 
selecting every 50th point from the chain to reduce the effects of correlations, and then for each point compute the distance to its 500th nearest neighbor (after downsampling). These distances should anti-correlate strongly with the local density.
For the nearest neighbor search and distance calculation, the parameter space is normalized by the 16-84\% percentile range along each dimension. We also perform ``reflection'' on $\sigma_{\log M}$ and $S_z$ dimensions at their lower boundaries in their prior to remove the boundary effect: each point in the chain is duplicated with the same parameters except for $\sigma_{\log M}$ which adopts the value of $0 - \sigma_{\log M}$ where 0 is the lower boundary of $\sigma_{\log M}$; subsequently the same procedure is also performed for $S_z$, so in the end we have 4 times the original number of points.

\begin{figure*}
   \includegraphics[width=0.97\textwidth]{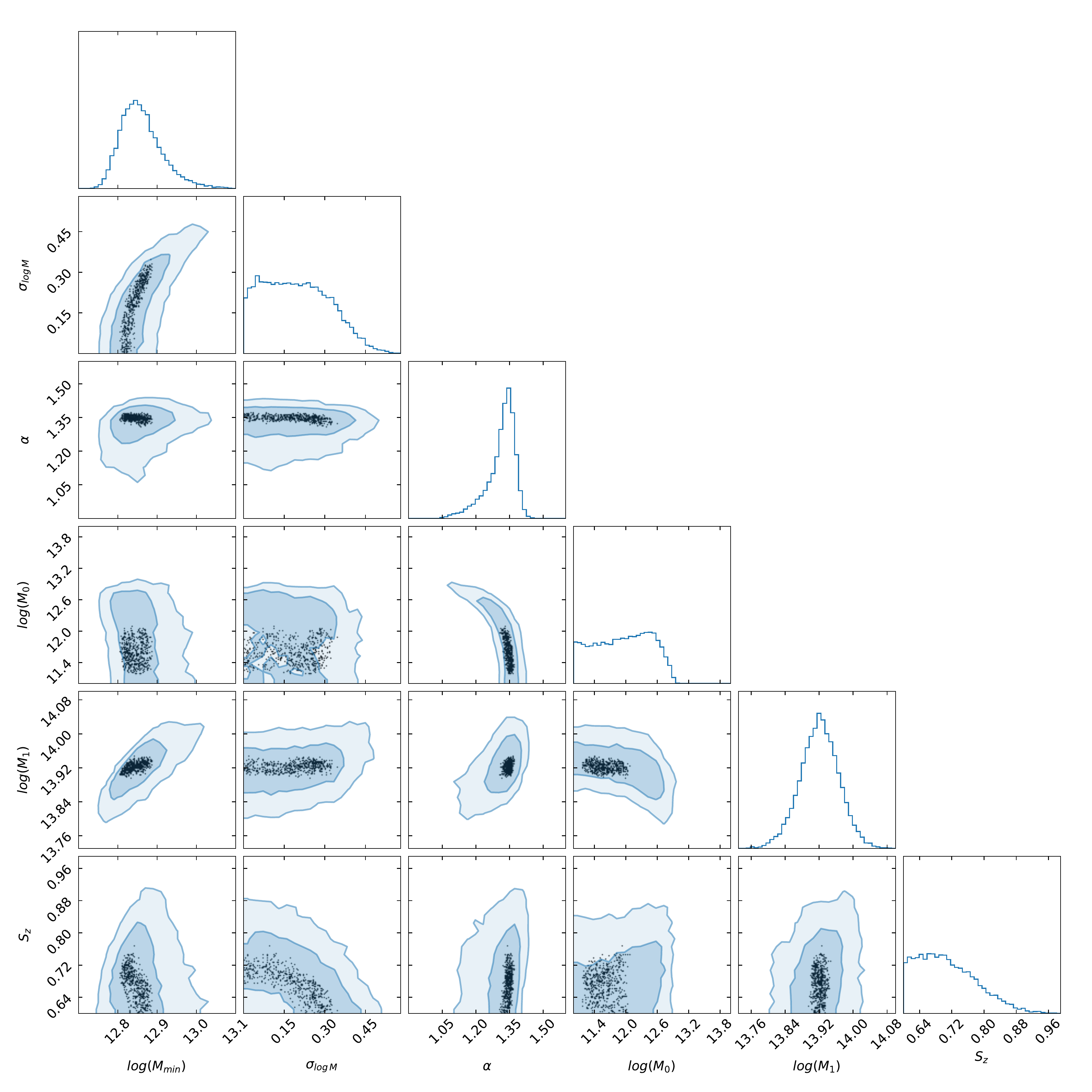}
   \centering
    \caption{The distribution of 500 points that are in the highest density region in parameter space. The contours of the full MCMC chain are also shown. The distribution of these points is much more compact than the distribution of points selected directly from MCMC likelihoods which is shown in the previous plot.}
    \label{fig:mcmc_highest_density}
\end{figure*}

We select the 500 points with the smallest neighbor distances. The distribution of these points are shown in Fig. \ref{fig:mcmc_highest_density}; note that these points have a much more compact distribution than those in Fig. \ref{fig:mcmc_highest_lilelihood}. For each of these points, we generate 100 HOD realizations and determine the $\chi^2$ for each. We then compute the averaged $\chi^2$ using the Hodges-Lehmann estimator \citep{hodges_estimates_1963}. Since there is still some scatter in the averaged $\chi^2$, we again select the 10 points with the smallest averaged $\chi^2$ and generate 1000 HOD realizations to get more accurate $\chi^2$ values. The point with the smallest averaged $\chi^2$ is selected as the set of best-fit parameters. In some redshift bins, the distributions of the 10 points are more compact than the 500 points; in other bins, the distributions are rather similar.

\section*{Data availability}
The LRG catalog and the photometric redshifts are derived from the publicly available Legacy Surveys imaging data\footnote{\url{https://www.legacysurvey.org/}}. The LRG catalog, masks, and the specific version of the photo-$z$'s used in the clustering analysis will be shared on reasonable request to the corresponding author.


\bibliographystyle{mnras}
\bibliography{LRG_clustering}


\bsp    
\label{lastpage}
\end{document}